\documentclass[11pt]{article}
\usepackage{amssymb,cite,amsmath,graphicx,epsf}
\usepackage{epstopdf}
\usepackage[totalwidth=450pt,totalheight=584pt]{geometry}
\newcommand\blank[1]{}
\newcommand{\fract}[2]{{\textstyle\frac{#1}{#2}}}

\newcommand{\eps}{\varepsilon}

\newcommand\ZZ{{\mathbb Z}}
\newcommand\RR{{\mathbb R}}
\newcommand\NN{{\mathbb N}}

\newcommand{\balpha}{\alpha\kern -6.7pt\alpha}
\newcommand{\bbalpha}{\alpha\kern -4.95pt\alpha}

\newcommand{\CN}{{\cal N}}

\newcommand{\CQ}{Q}
\newcommand{\CO}{{\cal O}}
\newcommand{\I}{A}

\newcommand\eq{\begin{equation}}
\newcommand\en{\end{equation}}
\newcommand\bea{\begin{eqnarray}}
\newcommand\eea{\end{eqnarray}}
\newcommand\nn{\nonumber}

\newcommand{\One}{{\hbox{{\rm 1{\hbox to 1.5pt{\hss\rm1}}}}}}
\renewcommand{\One}{{\mathbb 1}}
\renewcommand{\One}{{\rm 1\!\!1}}

\newcommand\ep{\epsilon}

\newcommand{\Ad}{\text{AdS}_5/\text{CFT}_4}
\newcommand{\so}{{*_{\bgammao}}}
\newcommand{\sx}{{*_{\bgammax}}}
\newcommand{\bgammax}{\bar{\gamma}_{\sf x}}
\newcommand{\bgammao}{\bar{\gamma}_{\sf o}}
\newcommand{\GammaX}{\Gamma_{\sf X}}
\newcommand{\GammaO}{\Gamma_{\sf O}}
\newcommand{\RRe}{\text{Re}}
\newcommand{\IIm}{\text{Im}}
\newcommand{\ba}{\begin{eqnarray}}
\newcommand{\ea}{\end{eqnarray}}
\newcommand{\vep}{\varepsilon}
\newcommand{\CG}{{\cal F}}
\newcommand{\be}{\begin{equation}}
\newcommand{\ee}{\end{equation}}
\newcommand{\CI}{I}
\newcommand{\CJ}{J}
\newcommand\ts{\text{d}}
\newcommand{\La}{\Lambda}
\newcommand{\s}{\tau}
\newcommand{\z}{s}
\newcommand{\zi}{t}
\newcommand{\gG}{{\rm g}}

\begin{document}

%GKV symbols:
\newcommand{\figQ}{{\includegraphics[scale=0.5]{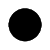}}}
\newcommand{\figyp}{{\includegraphics[scale=0.5]{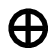}}}
\newcommand{\figym}{{\includegraphics[scale=0.5]{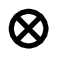}}}
\newcommand{\figw}{{\includegraphics[scale=0.5]{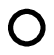}}}
\newcommand{\figv}{{\includegraphics[scale=0.5]{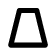}}}

%%%%%%%%%%%%%%%%%%%%%%%%%%%%%%%%%%%%%%%%%%%%%%%%%%%%%%%%%%%%%%%%%%%%%

\begin{titlepage}
\vskip 2.8cm
\begin{center}
{\Large {\bf Extended Y-system   for the  $\text{AdS}_5/\text{CFT}_4$ correspondence} } \\[5pt]
{\Large {\bf } }
\end{center}
\vskip 0.8cm

\centerline{Andrea Cavagli\`a%
\footnote{{\tt Andrea.Cavaglia.1@city.ac.uk}},
Davide Fioravanti%
\footnote{{\tt Fioravanti@bo.infn.it}}
and Roberto Tateo%
\footnote{{\tt Tateo@to.infn.it}}}
\vskip 0.9cm
\centerline{${}^{1,3}$\sl\small Dip.\ di Fisica Teorica
and INFN, Universit\`a di Torino,} \centerline{\sl\small Via P.\
Giuria 1, 10125 Torino, Italy}
\centerline{${}^{1}$\sl\small
Centre for Mathematical Science, City University London,}\centerline{\sl\small
Northampton Square, London EC1V 0HB, UK}
\centerline{${}^{2}$\sl\small INFN-Bologna and Dipartimento di Fisica,
Universit\`a di Bologna,} \centerline{\sl\small Via Irnerio 46, 40126 Bologna, Italy}
\vskip 1.25cm
\vskip 0.9cm
\begin{abstract}
\vskip0.15cm
\noindent
We study the analytic properties of the  $\text{AdS}_5/\text{CFT}_4$ Y functions. It is shown that  the  TBA  equations, including the dressing factor, can be obtained
from the Y-system with some additional information on the square-root discontinuities across semi-infinite
segments in the complex plane.  The Y-system extended by the  discontinuity relations constitutes
a  fundamental    set of local functional constraints that can be easily  transformed into integral form through Cauchy's
theorem.
\end{abstract}
\end{titlepage}
\setcounter{footnote}{0}
%%%%%%%%%%%%%%%%%%%%%%%%%%%%%%%%%%%%%%%%%%%%%%%%%%%%%%%%%%%%%%%%%%%%%
\newcommand{\resection}[1]{\setcounter{equation}{0}\section{#1}}
\newcommand{\appsection}[1]{\addtocounter{section}{1}
\setcounter{equation}{0} \section*{Appendix \Alph{section}~~#1}}
\renewcommand{\theequation}{\thesection.\arabic{equation}}
%\renewcommand{\theequation}{\arabic{equation}}
%%%%%%%%%%%%%%%%%%%%%%%%%%%%%%%%%%%%%%%%%%%%%%%%%%%%%%%%%%%%%%%%%%%%%
%%%%%%%%%%%%%%%%%%%%%%%%%%%%%%%%%%%%%%%%%%%%%%%%%%%%
\def\thefootnote{\fnsymbol{footnote}}
%
%%%%%%%%%%%%%%%%%%%%%%%%%%%%%%%%%%%%%%%%%%%%%%%%%%%%%%%%%%%%%%%%%%%%%
%%%  start of the paper  %%%%%%%%%%%%%%%%%%%%%%%%%%%%%%%%%%%%%%%%%%%%
%%%%%%%%%%%%%%%%%%%%%%%%%%%%%%%%%%%%%%%%%%%%%%%%%%%%%%%%%%%%%%%%%%%%%
%
\resection{Introduction and Summary}
The presence of a quantum two-dimensional integrable model (IM)~\cite{INT} in multicolor reggeised gluon scattering  was  discovered by Lipatov more than fifteen years ago~\cite{L}. However, only rather recently and thanks to the AdS/CFT correspondence~\cite{M-GKP-W}, the  connection between integrable models and supersymmetric gauge theories started to develop faster and faster, as we will very briefly summarise in the rest of this section.

For the purposes of the present paper, the relevant version of the AdS/CFT conjecture relates the free type IIB superstring theory on  the $\text{AdS}_5\times\text{S}^5$ curved space-time to the
planar limit of the  $\CN=4$ Super Yang-Mills theory (SYM) in four dimensions  living  on  the boundary of $\text {AdS}_5$~\cite{M-GKP-W}. The nature of the relation is a strong/weak coupling duality and thus  powerful to use, but difficult to test.

The 't Hooft planar limit is defined by the scaling of the colour number $N\rightarrow \infty$ and the SYM coupling $g_{YM} \rightarrow 0$ while keeping the coupling $Ng_{YM}^2=\lambda = 4 \pi^2 g^2$ finite\footnote{A warning about notations: another definition for the string coupling $g$ is also widely used, so that in many works is found the relation $\lambda = 16 \pi^2 g^2$. Appendix \ref{AF} can be consulted to match the different conventions.}. In this limit only planar Feynman diagrams survive~\cite{'thooft}. As part of the duality, once the string tension is set proportional to  $g$, the quantum energy of a specific string state corresponds to  the conformal dimension of  a local single-trace composite operator  $\CO$ in SYM. On the integrable model side, this dimension is an eigenvalue of the dilatation operator which is believed to be equivalent  to an integrable Hamiltonian. The coincidence of the one loop mixing matrix in the purely scalar sector with an integrable $so(6)$ spin chain was first proven~\cite{MZ} and then  extended to all the gauge theory sectors and at all loops in a way which shows integrability in a weaker sense, but still provides the investigators with many powerful tools (cf. \cite{BS} and references therein). More in detail, any composite operator $\CO$ can be thought of as a state of a hypothetical spin chain Hamiltonian, whose degrees of freedom  are the operators in the trace.

Although the actual form of the Hamiltonian describing the model at arbitrary values of the coupling  $g$  is still unknown, the large quantum number spectrum  turned out to be exactly described by certain Bethe Ansatz-like equations, which are thus called asymptotic Bethe Ansatz (ABA) equations (cf. \cite{BS,BES} and references therein).
The ABA leads to  the energy $E(g)$ which coincides with the anomalous dimension of a long operator:  $\Delta_{\CO}(g) = \Delta^{bare}_{\CO}+ E(g)$.

In integrable system language, the ABA equations
(Beisert-Staudacher's ~\cite{BS} with the final dressing phase~\cite{BES}) are the analogue of the
Bethe-Yang equations for scattering models describing the quantisation of momenta
for a finite number of interacting particles living on a ring with very large circumference~\cite{INT}, as
described in~\cite{Be-MM}.

In a parallel way, integrability in superstring theory was discovered at classical level~\cite{BPR},  extended to semiclassical and partially to quantum level in~\cite{KMMZ}. The string integrability investigations helped and were helped by the aforementioned Bethe Ansatz description, the paradigm of this interplay being the final conjectured form of the dressing phase~\cite{BES}.

As already mentioned, this is only part of the SYM/IM correspondence. In fact, an important limitation emerges as a consequence of the  asymptotic character of the Bethe Ansatz: the latter ought to be  modified by
finite size effects as soon as the site-to-site interaction range in the loop expansion of the dilatation operator becomes greater than the chain length. This wrapping  effect~\cite{Sieg:2005kd, AJK} is particularly relevant in the semiclassical string theory which covers the strong coupling regime though, for special reasons, it  may not effect  particular families of operators.

A first important step leading to the partial  solution of this problem   was made
by Bajnok and Janik in~\cite{Bajnok:2008bm}.
Adapting  the formulas  for the finite-size L\"uscher corrections~\cite{Luscher,KM}  to the $\Ad$ context, they were
able to predict   the four loop contribution to  the Konishi operator. The result was
readily  confirmed by the complicated diagrammatic calculations of~\cite{FSSZ}.
The method proposed by Bajnok and Janik   can  be extended  to higher  orders in $g^2$~\cite{Bajnok:2008qj,Bajnok:2009vm}
but -as in most of the known  perturbative schemes-    the technical complication increases very sharply with the loop order  and precise  non perturbative predictions are usually out of reach.

In 1+1-dimensional  massive relativistic scattering theories there is one well known way to treat finite-size  effects non perturbatively and exactly: the  Thermodynamic Bethe Ansatz (TBA) method. For this purpose, the TBA  was first proposed for the ground state energy by Al. B. Zamolodchikov
in~\cite{Zamolodchikov:1989cf} and adapted  to excited states in~\cite{Bazhanov:1996aq,Dorey:1996re}.

As a result  of the TBA procedure, the  exact finite-size energy can be written in terms of  the  pseudoenergies $\varepsilon_a$: the  solutions    of a system of non-linear  integral equations. Even for relatively simple relativistic systems, such as the sine-Gordon model, an exact and exhaustive study  of the TBA equations for excited states is an unfinished  business.

An alternative but equivalent  approach to excited states was adopted in~\cite{Fioravanti:1996rz, Feverati:1998va}.
Under the perspective of a different  non-linear integral equation, this idea was
applied to some sectors of the asymptotic Beisert-Staudacher equations in~\cite{FR, Freyhult:2007pz,BFR} and to the wrapping effects in  the Hubbard model~\cite{FFGR}. The latter  system is deeply related to the model studied  in this paper.

Starting from the mirror version of Beisert-Staudacher equations due to Arutyunov and Frolov~\cite{Arutyunov:2007tc} (see also~\cite{Arutyunov:2009zu}), the ground state TBA equations were recently and independently  proposed in~\cite{Bombardelli:2009ns, Gromov:2009bc, Arutyunov:2009ur}. The associated set of functional relations for the functions  $Y_a=e^{\varepsilon_a}$, the Y-system~\cite{Gaudin:1971, Takahashi:1971, Zamolodchikov:1991et, Kuniba:1992ev, Ravanini:1992fi},  was derived confirming  an earlier proposal by Gromov, Kazakov and Vieira coming from  symmetry arguments~\cite{Gromov:2009tv}.

\begin{figure}[h]
\centering
\includegraphics[width=6.5cm]{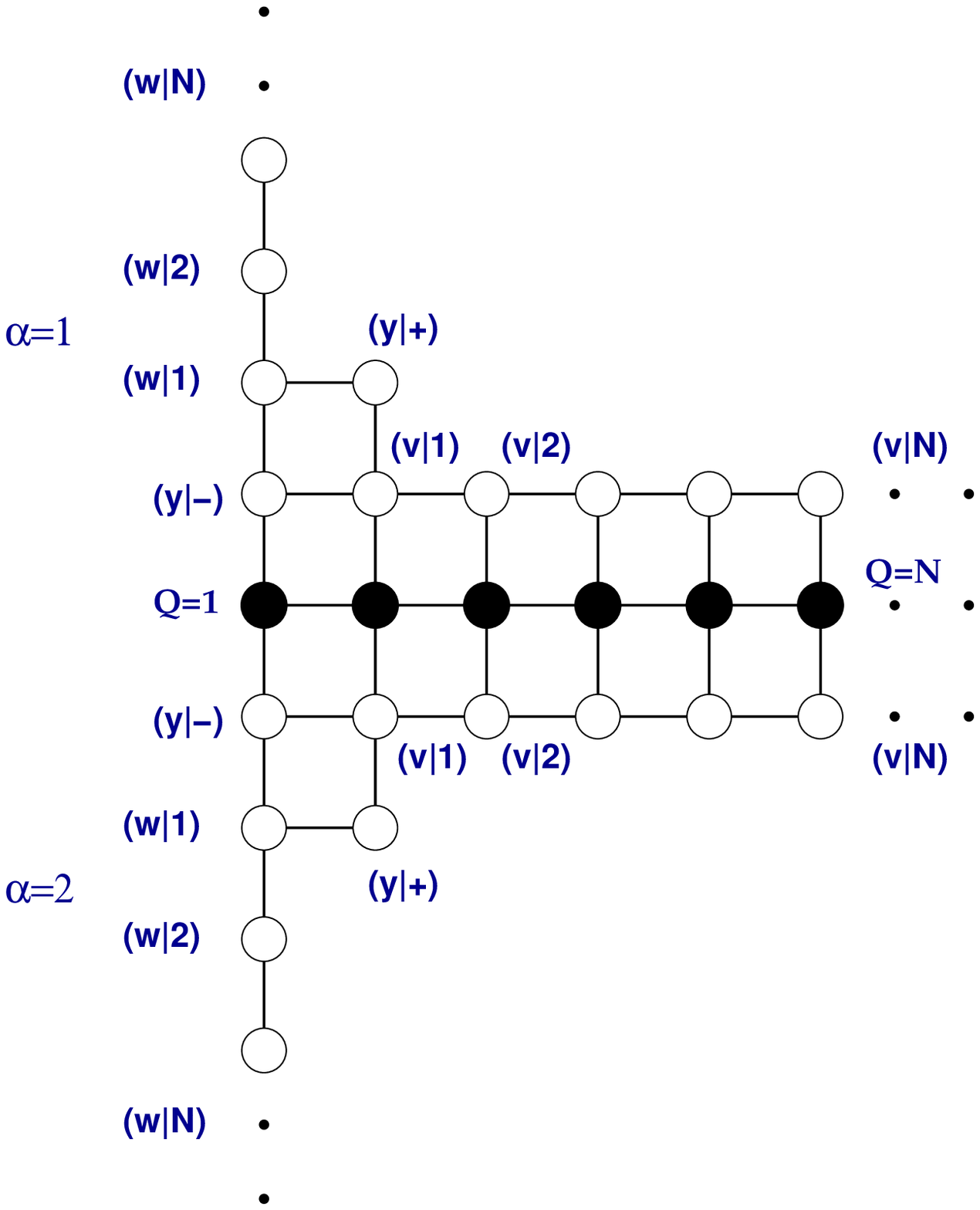}
\caption{ The Y-system  diagram   corresponding to the  $\Ad$  TBA equations. }
\label{N4LN}
\end{figure}

The $\Ad$ Y-system  conjectured in~\cite{Gromov:2009tv},  derived
in~\cite{Bombardelli:2009ns, Gromov:2009bc, Arutyunov:2009ur} and associated to  the
`T-hook' diagram represented in Figure~\ref{N4LN} is:
\eq
Y_{\CQ}(u-\fract{i}{g}) Y_{\CQ}(u+\fract{i}{g})= \prod_{\CQ'}\left(1+Y_{\CQ'}(u)\right)^{\I_{\CQ \CQ'}}
\prod_{\alpha}
{ \left(1+\frac{1}{Y_{(v|\CQ-1)}^{(\alpha)}(u)}\right)^{\delta_{\CQ,1}-1} \over {\left(1+\frac{1}{Y_{(y|-)}^{(\alpha)}(u)}\right)^{\delta_{\CQ,1}}}},
\label{eq:YQf}
\en
\eq
Y_{(y|-)}^{(\alpha)}(u+\fract{i}{g})  Y_{(y|-)}^{(\alpha)}(u-\fract{i}{g}) = \frac{\left(1+Y_{(v|1)}^{(\alpha)}(u)\right)}{\left(1+Y_{(w|1)}^{(\alpha)}(u)\right)}  \frac{1}{\left( 1+\frac{1}{Y_{1}(u)} \right)},\label{Yysystem}
\en
\eq
Y_{(w|M)}^{(\alpha)}(u + \fract{i}{g}) Y_{(w|M)}^{(\alpha)}(u-\fract{i}{g})=\prod_N\left(1+Y_{(w|N)}^{(\alpha)}(u)\right)^{\I_{MN}}
\left[\frac{\left(1+\frac{1}{Y^{(\alpha)}_{(y|-)}(u)}\right)}{\left(1+\frac{1}{Y^{(\alpha)}_{(y|+)}(u)}\right)}\right]^{\delta_{M, 1}},\label{yw}
\en
\eq
Y_{(v|M)}^{(\alpha)}(u+\fract{i}{g}) Y_{(v|M)}^{(\alpha)}(u-\fract{i}{g})=\frac{\prod_N\left(1+Y_{(v|N)}^{(\alpha)}(u)\right)^{\I_{MN}}}{\left(1+\frac{1}{Y_{M+1}(u)}\right)}
\left[\frac{\left(1+Y^{(\alpha)}_{(y|-)}(u)\right)}{\left(1+Y^{(\alpha)}_{(y|+)}(u)\right)}\right]^{\delta_{M, 1}},
\label{yv}
\en
where $\I_{1,M}=\delta_{2, M}$, $\I_{NM}= \delta_{M, N+1}+\delta_{M, N-1}$  and  $\I_{MN}=\I_{NM}$.

In the integrable model framework  the  Y-systems play a very central r\^ole.
Firstly, a Y-system  exhibits a very   high degree of universality. Not only the whole set of excited states of a given theory is associated  to a single  Y-system but also many   different models  may have  identical   Y-systems.
Two excited states of the same theory or   two states of different  models sharing  a common Y-system    differ in the   analytic properties of the Y functions inside a fundamental strip of the  complex rapidity plane. Given this analytic information the Y-system can be  easily transformed to the integral  TBA form. Roughly speaking, two different models have  different leading asymptotic behaviors, while  different states of the same model  differ in the  number and positions  of the $1+Y_a$ zeros in the fundamental strip. In relativistic models the Y functions are in general  meromorphic  in the rapidity $u$ with zeros and poles both linked to    $1+Y_a$ zeros through the Y-system.

For the ground state energy numerics is  reliable and the accuracy is in general  very high. However, for excited states the situation is  complicated by the presence  of the finite number of auxiliary equations constraining the positions of this  special  subset of zeros. Unfortunately,  both the number of special  zeros and their positions in the complex rapidity plane can change drastically as the coupling constant or the system size   interpolate  between the  infrared and the ultraviolet  regimes. Therefore,  the situation at moderate
$L$  may be substantially different from the infrared distribution described by the asymptotic Bethe-Yang equations~\cite{Bazhanov:1996aq,Dorey:1996re, Dorey:1997rb}.

The situation for the $\Ad$-related model is further complicated by the presence of square root branch discontinuities inside and at the border of the fundamental strip $|\IIm(u)|\le 1/g$. According to the known    Y$\rightarrow$ TBA transformation procedures this extra information
should be independently supplied. However,  it was pointed out in~\cite{Arutyunov:2009ur} (see also~\cite{Gromov:2009tq}) that such discontinuity information
is stored into   functions  which depend non locally on the TBA pseudoenergies. In other words, they  crucially depend on the   particular  excited state under consideration.

The main objective  of this paper is to show  that this problem can be overcome and in particular that the discontinuity information is encoded in the Y-system together with the following  set  of local and  state-independent   functional relations. Setting
\eq
\Delta(u)= \left[\ln Y_1(u)\right]_{+1},
\label{delta0}
\en
then $\Delta$ is the function introduced in~\cite{Arutyunov:2009ur} and the local discontinuity relations are:
\eq
\left[
\Delta\right]_{\pm 2N}=\mp\sum_{\alpha=1,2} {\biggl(} \biggl[\ln\biggl(1+ \frac{1}{Y_{(y|\mp)}^{(\alpha)}}\biggr)\biggr]_{\pm 2N}+\sum_{M=1}^{N} \biggl[\ln\biggl(1+{1 \over Y^{(\alpha)}_{(v|M)}}\biggr)\biggr]_{\pm (2N-M)}
+\ln{\biggl({Y_{(y|-)}^{(\alpha)} \over Y_{(y|+)}^{(\alpha)}}\biggl)}\biggl),\label{d2}
\en
\eq
\biggl[\ln\biggl(\frac{Y^{(\alpha)}_{(y|-)}}{Y^{(\alpha)}_{(y|+)}}\biggl)\biggr]_{ \pm 2N}=- \sum_{\CQ=1}^N\left[\ln\left(1+{1 \over Y_{\CQ}}\right)\right]_{\pm (2N-\CQ)},
\label{d1}
\en
with $N=1,2,\dots,\infty$  and
\eq
\left[\ln{Y^{(\alpha)}_{(w|1)}} \right]_{\pm 1}= \ln\biggl( { 1+1/Y_{(y|-)}^{(\alpha)} \over  1+1/Y_{(y|+)}^{(\alpha)}}\biggl),~~~~
\left[\ln{Y^{(\alpha)}_{(v|1)}} \right]_{\pm 1}=
\ln\biggl( { 1+Y_{(y|-)}^{(\alpha)} \over  1+Y_{(y|+)}^{(\alpha)}}\biggl),
\label{d3}
\en

where the  symbol $[f]_Z$  with $Z \in \ZZ$ denotes  the discontinuity of $f(z)$
\eq
\left[ f \right]_{Z} =  \lim_{\ep \rightarrow 0^+} f(u+i Z/g+i \ep)- f(u+iZ/g -i \ep),
\label{dis0}
\en
on the  semi-infinite segments described by
$z=u+ i Z/g$ with  $u \in (-\infty,-2) \cup (2,+\infty)$ and the function
$\left[ f \right(u)]_{Z}$ is the analytic extension of the discontinuity  (\ref{dis0}) to generic complex values of $u$. To retrieve the TBA equations, the extended Y-system has to be supplemented with the asymptotics
\bea
e^{\Delta(u)}&\sim u^{2L}~~~\text{for}~ u\rightarrow \infty,~\IIm(u)<0;~~~~~e^{\Delta(u)}&\sim 1/u^{2L} ~~~\text{for}~ u\rightarrow \infty,~\IIm(u)>0,~~~
\label{d4}
\eea
which capture all the dependence on the scale $L$.

In this paper, instead of describing how  relations (\ref{d2}-\ref{d3})   can be  deduced   from the TBA equations,  we will show   how  the ground state TBA equations for both the mirror and the direct  $\Ad$ theories can be derived from (\ref{d2}-\ref{d3})  and the Y-system using Cauchy's integral theorem. Many other  interesting results will emerge along the way on the analytic properties of the Y functions, on  the   dressing factor and on the quantisation of $L$.

The rest of this paper is organised as follows. Section~\ref{TBAM} contains
the TBA equations  of~\cite{Bombardelli:2009ns, Gromov:2009bc, Arutyunov:2009ur} written in a form more  appropriate to the study of their analytic structure. A previously unnoticed link between the quantisation of the total momentum, the  dilogarithm trick~\cite{Zamolodchikov:1989cf} and  the trace condition  is discussed  in   Section~\ref{dilog}.

The $\Ad$-related Y-system is described  in Section~\ref{Ysystem}; its  validity as the rapidity parameter  $u$ is moved in the complex plane is briefly discussed together with some preliminary comments on the analytic structure of the Y functions. The functional relations
versus TBA  transformation method   for a particular equation involving the fermionic $y$-particles  is described in  detail in Section~\ref{sectionDis}. The derivation of the discontinuity function $\Delta$ from the local Y-system and the relevant discontinuity
relations is given in Section~\ref{deltas}. The TBA equations  for the  $w-$, $v-$ and $\CQ$-particles are derived   in
Sections~\ref{wnodes},\ref{vnodes} and \ref{Qnodes}, respectively.

As a preliminary  application of the   method, we have derived the TBA for the direct theory: this result is described in Section~\ref{TBADD}. Finally Section~\ref{conclusions} contains our conclusions.
There are  also seven appendices. The S-matrix elements involved in the definition of the TBA kernels are given in Appendix~\ref{AA}. Appendix~\ref{Akernels} presents some previously known identities with an emphasis on aspects of analytic continuation. Appendices~\ref{AB} to \ref{AE}  contain  the most technical parts of the calculations. In particular,
in Appendices~\ref{AB} and \ref{AC}  we prove the equivalence  between the two different explicit  expressions for the  mirror   dressing factor $\Sigma$  and  $\hat{\sigma}$ given in~\cite{Arutyunov:dressingfactor}  and~\cite{Gromov:2009bc}, respectively.
Different conventions  have been adopted in~\cite{Bombardelli:2009ns, Gromov:2009bc, Arutyunov:2009ur} for the string coupling, the labeling and  the definition of the Y functions and the TBA kernels: the purpose of  Appendix \ref{AF} is to provide the reader with a concise dictionary.
%
%%%%%%%%%%%%%%%%%%%%%%%%%%%%%%%%%%%%%%%%%%%%%%%%%%%%%%%%%%%%%
%
\resection{The TBA equations}
\label{TBAM}
The origin of the non-standard properties of the $\Ad$ thermodynamic Bethe Ansatz  system rests on the
presence, in the S-matrix elements,  of a  double-valued function $x(u)$  defined as
\eq
x(u)= \left( \frac{u}{2} -i \sqrt{ 1- \frac{u^2}{4}} \right).~
\label{xu}
\en
In the first Riemann sheet:   $\IIm(x)<0$   and $x$  behaves   under complex conjugation as  $x(u)=1/x^*(u^*)$.
It is convenient  to parameterize the TBA pseudoenergies  in terms of a  common rapidity variable $u$, analogously the kernels will be defined in terms of a pair of complex variables~\cite{Bombardelli:2009ns, Gromov:2009bc, Arutyunov:2009ur}.  The relationship between the Hubbard variables~\cite{Takahashi,onedHubbard} and $u$  is~\cite{Arutyunov:2009zu}:
\begin{align}
\label{pt}
\begin{split}
&\tilde{p}^{\CQ}(u) = \frac{i g}{2} \left( \sqrt{4 - \left(u+ i \fract{\CQ}{g}\right)^2}
-\sqrt{4 - \left(u- i \fract{\CQ}{g}\right)^2} \right), \\
&i\,e^{-iq} = y(u),~~~~~~\lambda = u=2 \sin(q).
\end{split}
\end{align}
The   double-valued function $y(u)$ can be written in terms of $x(u)$ as
\eq
y(u) =  \left\{ \begin{array}{lll}
 x(u) & \hbox{for} & \IIm(y)<0;  \\
& & \\
1/x(u) & \hbox{for} & \IIm(y)>0.
\end{array}\right.
\label{yy}
\en
The  thermodynamic Bethe Ansatz  equations derived in~\cite{Bombardelli:2009ns, Gromov:2009bc, Arutyunov:2009ur} and
the associated Y-system can be
encoded in the diagram represented in Figure~\ref{N4LN} and originally proposed in~\cite{Gromov:2009tv}. The reader is addressed to~\cite{ZoliRev} for a  concise up-to-date  summary of the main results on TBA and excited states.
The TBA equations, at arbitrary chemical potentials~\cite{Klassen:1990dx, Martins:1991hw,Fendley:1991xn},  are:
\begin{align}
\vep_{\CQ}(u)=\mu_{\CQ} +&L\,\tilde{E}_{\CQ}(u)
-\sum_{\CQ' } L_{\CQ'}*\phi^{\Sigma}_{\CQ' \CQ}(u)
+ \sum_{\alpha} {\Big (}\sum_{M} L^{(\alpha)}_{v|M}*\phi_{(v|M),\CQ}(u)
+ L^{(\alpha)}_{y}\so \; \phi_{y,\CQ}(u) {\Big )},\label{TBA1} \\
\vep_{y}^{(\alpha)}(u)&=\mu_{y}^{(\alpha)}-\sum_{\CQ}  L_{\CQ}*\phi_{\CQ,y}(u)
+ \sum_{M}  (L^{(\alpha)}_{(v|M)}-L^{(\alpha)}_{(w|M)}  )*\phi_{M}(u), \label{TBA2}\\
\vep^{(\alpha)}_{(w|K)}(u)&=\mu_{(w|K)}^{(\alpha)}+\sum_{M} L^{(\alpha)}_{(w|M)}*\phi_{M K}(u)+ L^{(\alpha)}_{y}\so \; \phi_{K}(u)
,
\label{TBA4} \\
\vep^{(\alpha)}_{(v|K)}(u)&=\mu_{(v|K)}^{(\alpha)}-
\sum_{\CQ} L_{\CQ}*\phi_{\CQ,(v|K)}(u)
+\sum_{M} L^{(\alpha)}_{(v|M)}*\phi_{M K}(u)+ L^{(\alpha)}_{y}\so \;\phi_K(u)
,\label{TBA3}
\end{align}
where $\alpha=1,2$, $K=1,2,\dots$,
\eq
\tilde{E}_{\CQ}(u)= \ln{\frac{x(u-i\CQ/g)}{x(u+i\CQ/g)}}~
\en
and the  inverse of the  temperature  $L \in \NN$ is a discrete SYM variable.

In the following we will agree that, when not specified otherwise, the sums $\sum_{\alpha}$, $\sum_{\s}$, $\sum_K$ (for any capital letter $K$) will always   stand  for $\sum_{{\alpha}=1,2}$, $\sum_{{\s}=\pm1}$ and $\sum_{K=1}^{\infty}$, respectively.
Further,
\eq
Y_a(u)=e^{\vep_a(u)},~~L_a(u)= \ln(1+1/Y_a(u)),~~\La_a(u)=\ln(1+Y_a(u)),
\label{LA}
\en
for any choice of the collective index $a$ and  $\{ \mu_a \}$ will be the set of chemical potentials.
The symbols `$*$' and `$*_{\gamma}$' denote respectively   the convolutions
\eq
\CG*\phi(u)=\int_{\RR} dz \; \CG(z) \,\phi(z,u),~~\CG *_{\gamma} \phi(u)=\oint_{\gamma} dz \; \CG(z) \,\phi(z,u).
\en

\begin{figure}
\begin{minipage}[b]{0.5\linewidth}
\centering
\includegraphics[width=7cm]{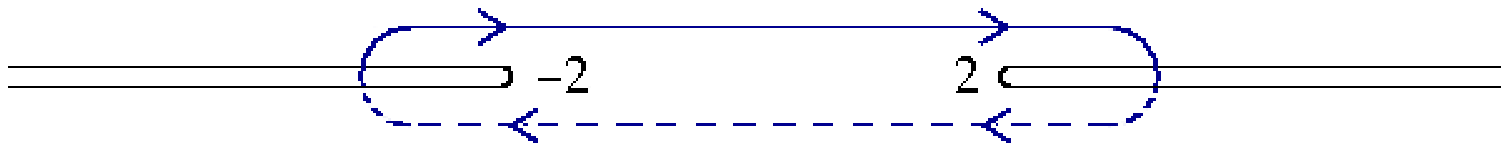}
\caption{The contour $\bgammao$.}
\label{gM}
\end{minipage}
\hspace{0.3cm}
\begin{minipage}[b]{0.5\linewidth}
\centering
\includegraphics[width=7cm]{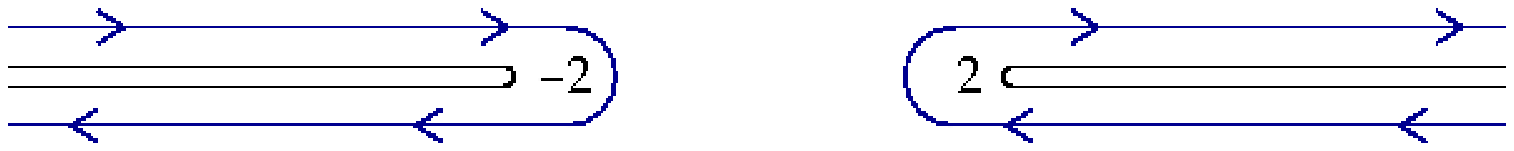}
\caption{ The contour $\bgammax$.}
\label{gD}
\end{minipage}
\end{figure}

The kernels are
\eq
\phi_{ab}(z,u)= {1 \over 2 \pi i} \frac{ d}{dz} \ln S_{ab}(z,u),
\en
where  the `S-matrix' elements $S_{ab}$ are listed in Appendix~\ref{AA}.\\
\noindent The kernel $\phi^{\Sigma}_{\CQ'\CQ}$ appearing in
equation  (\ref{TBA1}) can be naturally separated  into two parts
\eq
\phi^{\Sigma}_{\CQ'\CQ}(z,u) = - \phi_{\CQ'\CQ}(z-u) - 2 K^{\Sigma}_{\CQ'\CQ}(z,u),
\en
with
\eq
\phi_{\CQ' \CQ}(u)= {1 \over 2 \pi i} \frac{d}{du} \ln S_{\CQ'\CQ}(u),~~K^{\Sigma}_{\CQ'\CQ}(z,u)={1 \over 2 \pi i} \frac{d}{dz} \ln \Sigma_{\CQ'\CQ}(z,u).
\label{pqq}
\en
In (\ref{pqq}), $\Sigma_{\CQ'\CQ}$ is the improved  dressing factor~\cite{Dorey:2007xn} evaluated in the mirror kinematics~\cite{Arutyunov:dressingfactor}.
Starting form the explicit expression for  $\Sigma_{\CQ'\CQ}$  given ~\cite{Arutyunov:dressingfactor}, a main result of Appendices~\ref{AB} and \ref{AC}  is the
following compact expression for  the kernel $K^{\Sigma}_{\CQ'\CQ}$, valid under convolution with $\sum_{\CQ'} L_{\CQ'}$ (see also the end of Section~\ref{Qnodes}):
\eq
K^{\Sigma}_{\CQ'\CQ}(z,u) =
\frac{1}{2 \pi i}\frac{d}{dz}{\ln\Sigma}_{\CQ'\CQ}(z, u) =\oint_{\bgammax} ds \; \phi_{\CQ', y}(z, s) \oint_{\bgammax} dt \; K_{\Gamma}^{\left[2\right]}(s- t)\phi_{y, \CQ}(t, u),
\label{eq:hofman2maintext}
\en
where the contour $\bgammax$ is represented in Figure~\ref{gD} and
\eq
K_{\Gamma}^{[N]}(u-z) = {1 \over 2 \pi i } {d \over du} \ln {\Gamma(N/2-ig(u-z)/2)  \over \Gamma(N/2+ig(u-z)/2) }.
\label{gammaN}
\en
The result (\ref{eq:hofman2maintext}) shows  that  $\Sigma_{\CQ'\CQ} \equiv \hat{\sigma}_{\CQ'\CQ}$, where $\hat{\sigma}$ is the mirror  dressing factor  of~\cite{Gromov:2009bc,Volin:2009uv, Volin:2010cq}.

It is interesting that the  result (\ref{eq:hofman2maintext}) has precisely the double convolution  form predicted  by~\cite{Janik:2008hs}
and formally identical to that of the dressing factor in the  direct channel~\cite{Dorey:2007xn}.
Further, the equilibrium free energy at finite temperature $T=1/L$ is
\eq
 \tilde{F}(L)=-{1 \over L} \sum_{\CQ} \int_{\RR} {du \over 2 \pi}  \,  {d\tilde{p}^{\CQ} \over du}  L_{\CQ}(u),
\label{FL}
\en
with
\eq
\tilde{p}^{\CQ}(u)=g x(u - i \CQ/g) -g x(u + i \CQ/g) + i \CQ.
\en
We shall  restrict the set of allowed  chemical potentials by
\eq
\prod_{\CQ'} e^{ \mu_{\CQ'} \; C_{\CQ' \CQ} }=e^{2\mu_y^{(\alpha)} - \mu_{(v|1)}^{(\alpha)}+\mu_{(w|1)}^{(\alpha)}}=\prod_{M} e^{ \mu_{(w|M)}^{(\alpha)} \;C_{MK} }=\prod_{M} e^{\mu_{(v|M)}^{(\alpha)} \;C_{MK} } =
1,
\label{constmu}
\en
with  $C_{MN}= 2 \delta_{M,N} - \I_{MN}$,  $\I_{1,M}=\delta_{2, M}$, $\I_{NM}= \delta_{M, N+1}+\delta_{M, N-1}$  and  $\I_{MN}=\I_{NM}$.
Then in the derivation of the Y-system  the chemical potentials cancel and in all the  subsequent considerations they can be ignored and eventually restored in the final equations. The case with  all vanishing chemical
potentials  apart for  $\mu_y^{(1)}= -\mu_y^{(2)}=i \pi$ is perhaps the most interesting: it corresponds to the calculation of the Witten
index~\cite{Cecotti:1992qh} and,  through the relation
\eq
E_0(L)= L \tilde{F}(L),
\en
to the zero energy protected ground state of  $\Ad$.
A vanishing ground state energy, or in general  any protected state, should  correspond to a singularity in
the associated  TBA solution. A  way  to explore the neighborhoods of this point, while preserving the Y-systems structure,  is to
slightly displace  the chemical potentials from these critical values such that  (\ref{constmu}) are still fulfilled.

As we shall see in more details in the following sections,  the functions  $Y_a(u)=e^{\vep_a(u)}$ solutions of the TBA equations  (\ref{TBA1}-\ref{TBA3}) live on
multi-sheeted coverings of the complex plane with an infinite number of square-root branch points
 in the set $u \in \{ \pm 2+ i m/g \}$ with $m \in \ZZ$. For the mirror-$\Ad$   theory under consideration   all the cuts are
conventionally set  parallel to the real axis and external to the strip $|\RRe (u)|<2$.

\begin{figure}[h]
\centering
\includegraphics[width=11cm]{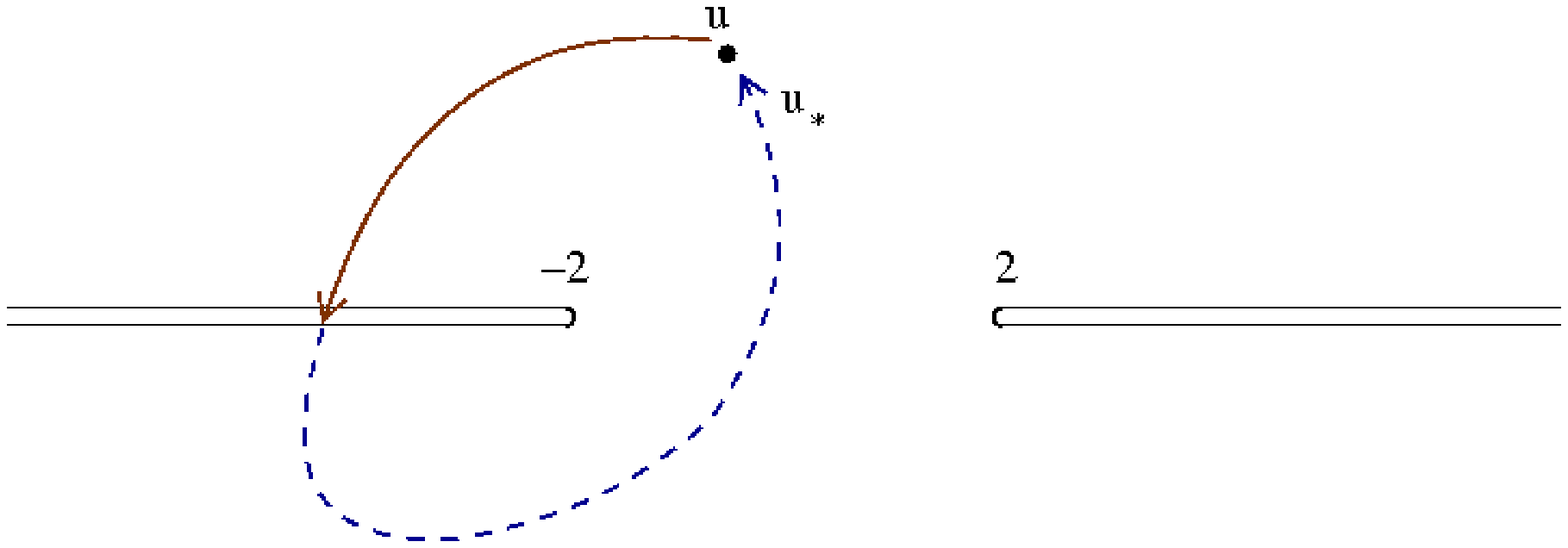}
\caption{ The second sheet image $u_*$ of $u$. }
\label{cutfinal}
\end{figure}

In particular, as a direct consequence of  the mapping between the function  $Y^{(\alpha)}_y(q)$ of~\cite{Bombardelli:2009ns}
and $Y^{(\alpha)}_y(u)$ through  the relation  $i\,e^{-iq} = y(u)$ and (\ref{yy}),
$Y^{(\alpha)}_y(u)$ has   a pair of square-root branch points on the real axis  at $u=\pm 2$. In equations
(\ref{TBA1}-\ref{TBA3}), the
integration contour $\bgammao$, represented in   Figure~\ref{gM},
is a negative oriented closed path surrounding  precisely this  pair of  singularities.
In the  following sections we shall denote with $Y^{(\alpha)}_{(y|-)}(u)$,  or when it will not be  possible source of  confusion simply  by $Y^{(\alpha)}_y(u)$, the first sheet evaluation of  $Y^{(\alpha)}_y$  and with $Y^{(\alpha)}_{(y|+)}(u)=Y^{(\alpha)}_y(u_*)$ the  evaluation  obtained by replacing  $u$ with  its  second sheet image  $u_*$ reached by  analytic continuation through the branch cut  $u \in (-\infty,-2)$ (see, Figure~\ref{cutfinal}). For  the function $x$ defined in  (\ref{xu}) and appearing in the definition of the S-matrix elements   we have
\eq
x_+(u)=x(u_*)=1/x(u),~~\IIm(x(u_*))>0.
\en
\resection{Total momentum quantisation,  the  trace condition and the dilogarithm trick}
\label{dilog}
The study of excited states goes beyond the objectives  of this paper. Here, we simply   remind to the reader that in analogy with the early results for relativistic models~\cite{Bazhanov:1996aq,Dorey:1996re} excited state TBA equations  can be obtained from (\ref{TBA1}-\ref{TBA3}) by  replacing the contour
of integration with a more general complex path $\gamma_a$ enclosing  a finite  number of  poles   $\{u_a^{(k)} \}$ of  ${  dL_a(u) \over du }$. The number of active  singularities, or equivalently of extra residues $\ln S_{ab}(u_a^{(k)},u)$  explicitly
appearing in the excited state TBA variants,  crucially depends on the particular  energy level $E_n(L)$ under consideration, the scale $L$ and the coupling. The first  simple examples of active singularity transitions
were  observed in~\cite{Bazhanov:1996aq,Dorey:1996re}  and the complete  range of possibilities discussed in~\cite{Dorey:1997rb}. Adopting these ideas, excited state TBA equations for  $\Ad$  were  proposed  in~\cite{Gromov:2009bc, Arutyunov:2009ax}
and partially studied in~\cite{Gromov:2009zb,Arutyunov:2009ax, Arutyunov:2010gb, Balog:2010xa, Frolov:2010wt}.

Finally, we would like to remark   that the quantity
\eq
P^{(n)}=\sum_Q \int_{\gamma^{(n)}_Q}  {du \over 2 \pi}
{d  \tilde E_Q(u) \over du} L_Q(u) =  \sum_Q \left(\int_{\RR} {du \over 2 \pi }
{d  \tilde E_Q(u) \over du} L_Q(u) + \sum_k \left(r_Q^{(k)} p^Q(u_Q^{(k)}) \right) \right),
\label{Pn}
\en
can be computed exactly adapting  the standard dilogarithm trick (see, for example~\cite{Zamolodchikov:1989cf}).
In equation (\ref{Pn}),  $r_Q^{(k)}=+1$ or $-1$ is the residue of  ${  dL_Q(u) \over du }$ at  $u=u_Q^{(k)}$ and $p^Q \equiv i {\tilde E}_Q$ is the single particle momentum in the direct theory.
Considering the possible monodromy properties of the Rogers dilogarithm one can argue that
\eq
P^{(n)}= {2 \pi Z^{(n)} \over L},~~Z^{(n)} \in \ZZ,
\label{Pn1}
\en
where $Z^{(n)}=0$ for the ground state and for any zero momentum  state. Therefore, we have found in $P^{(n)}$ a quantity which is naturally quantised  and it   is a very natural candidate  to be the  total momentum of the direct theory.
To our knowledge, even in the context of excited-state  TBA equations for relativistic models, this fact was never noticed  before.
Then, it is natural to expect that the momentum condition  (trace condition) on the physical states should correspond to the stronger requirement $Z^{(n)}/L \in \ZZ$. The latter  constraint  was first  imposed in~\cite{Gromov:2009bc, Gromov:2009at} and used to infer the equivalence between the TBA  equations in~\cite{Bombardelli:2009ns, Arutyunov:2009ur, Arutyunov:2009ax} and~\cite{Bombardelli:2009xz}
with  those in~\cite{Gromov:2009bc} and~\cite{Gromov:2009at}, respectively (see also  the related discussion in~\cite{ZoliRev}, and the end of Appendix \ref{AF}). Although this is a very interesting observation, we would like to underline here  that for a generic  physical state with non-zero momentum  a  difference $\Delta\mu_a=i2 \pi k_a$ with $k_a \in \ZZ$ between  the two sets of chemical potentials would still remain. As  it is well known,  the energy levels    $E_n(L)$ are in general   not $2\pi i$-periodic in the $\mu_a$'s and this  very  small difference  between  the two sets of  TBA equations  may  be  not   totally harmless.
%
%
%%%%%%%%%%%%%%%%%%%%%%%%%%%%%%%%%%%%%%%%%%%%%%%%%%%%%%%%%%%%%%%%%%%%%%%%%55
%
%
%%%%%%%%%%%%%%%%%%%%%%%%%%%%%%%%%%%%%%%%%%%%%%%%%%%%%%%%%%%%%%%%%%%%
%
\resection{The Y-system for the AdS/CFT correspondence}
\label{Ysystem}
A consequence of the non-trivial analytic properties of the $\Ad$ Y functions is that the product  $Y_a(u-i/g ) Y_a(u+i/g)$
requires a prescription on how to pass from $(u-i/g)$ to $(u+i/g)$.
If $Y_a(u+i/g)$ can be obtained from $Y_a(u-i/g)$ by shifting vertically  $u \rightarrow u+i 2/g$ for $\RRe(u) \in  (-2, 2)$, in general this is not true for  $u$  outside this region. Only  applying the two shifts
$\pm i /g$ to $Y_a(u)$ starting from $\RRe(u) \in (-2, 2)$  leads to the local Y-system of equations (\ref{eq:YQf}-\ref{yv}) above. A very direct way of deriving the Y-system is simply by analytic continuation of the TBA equations. The relevant identities involving the TBA kernels are collected in Appendix~\ref{Akernels}; here is a brief summary of the derivation:
\begin{itemize}
\item{$\CQ$-particles: The Y-system equation (\ref{eq:YQf})   is obtained from the TBA equation (\ref{TBA1}) using properties
(\ref{eq:nrev3}) and  (\ref{asy}).
}
\item{$y$-particles:
The Y-system equation (\ref{Yysystem}) for  $Y^{(\alpha)}_{(y|-)}(u) \equiv Y^{(\alpha)}_y(u)$ is obtained from the TBA equation (\ref{TBA2}) using
properties (\ref{eq:propkernela}) and  (\ref{eq:nrev}).
}
\item{$w$-particles:
The Y-system equation (\ref{yw})  is obtained from the TBA equation  (\ref{TBA4}) using properties (\ref{eq:propkernela}) and (\ref{eq:nrev}).
}
\item{ $v$-particles: Starting  from the TBA equation (\ref{TBA3})  and using properties (\ref{eq:propkernela}), (\ref{eq:nrev}) and     (\ref{eq:nrev2}) we have:
\begin{align}
Y_{(v|M)}^{(\alpha)}(u+\fract{i}{g})  Y_{(v|M)}^{(\alpha)}(u-\fract{i}{g})=&
\frac{\prod_N\left(1+Y_{(v|N)}^{(\alpha)}(u)\right)^{\I_{MN}}}{\left(1+\frac{1}{Y_{M+1}(u)}\right)}
\left[\frac{\left(1+\frac{1}{Y_{(y|-)}^{(\alpha)}(u)}\right)}{\left(1+\frac{1}{Y^{(\alpha)}_{(y|+)}(u)}\right)}\right]^{\delta_{M, 1}} \label{ymv}\\
&\times \exp  \left({\delta_{M,1} \sum_{\CQ}  \left(L_{\CQ}*\phi_{\CQ,(y|-)}(u) -L_{\CQ}*\phi_{\CQ,(y|+)}(u)  \right) } \right). \nn
\end{align}
Now, recall that we know a priori that the relevant values of the functions $Y^{(\alpha)}_y$ are from  two different sheets, which are connected by encircling one of the branch points at $\pm 2$. The difference between these two determinations of $\ln Y_y^{(\alpha)}$ can be easily read from the TBA equation (\ref{TBA2}) to coincide with the integral expression on the rhs of (\ref{ymv}):
\eq
\ln\frac{Y^{(\alpha)}_{(y|-)}(u)}{Y^{(\alpha)}_{(y|+)}(u)} =  \left[\ln Y^{(\alpha)}_{y}(u)\right]_{0}
=-\sum_{\CQ} \left(L_{\CQ}*\phi_{\CQ,(y|-)}(u)- L_{\CQ}*\phi_{\CQ,(y|+)}(u) \right).
\label{eq:disctilde}
\en
Equations (\ref{eq:disctilde}) and  (\ref{ymv}) then lead to (\ref{yv}).
}
\end{itemize}
Apart for some minor subtleties, equations (\ref{eq:YQf}-\ref{yv})  derived in this way match  the
proposal~\cite{Gromov:2009tv} by  Gromov, Kazakov and Vieira.

Finally, we would like to mention that although  the Y-system  was derived
for  $|\RRe(u)|<2$, the  coupled  equations (\ref{eq:YQf}-\ref{yv}) are simultaneously valid for any complex value of  $u$. To see this we can adapt to the current situation the argument first used in a similar context in~\cite{Dorey:1997rb}.
Subtracting the right-hand sides of (\ref{eq:YQf}-\ref{yv})  from the left-hand sides yields a set of  analytic functions (depending on the   same variable  $u$) which are  identically zero. These functions therefore
remain zero during any process of analytic continuation, and so the Y-system
and  its consequences hold equally for all complex values of $u$.

Table~\ref{table1} shows the location of the square branch points for the various Y's. For any analytic continuation path connecting the region $|\RRe(u)|<2$ to a generic point $u'$, this pattern  of singularities dictates unambiguously which branch of every Y function should be selected in order for the Y-system to hold. Notice that the most natural choice is to avoid crossing any of the semi-infinite segments $(-\infty, -2) \cup (2, +\infty)+im/g$, $m\in \ZZ$, so that  all the Y functions are simultaneously evaluated on the same sheet. Because it contains the physical values of the Y's, this will be referred to as the first Riemann sheet.

A short
discussion  on the independent analytic continuation of  $Y_1$  around the branch points at $u=-2 \pm i/g$  and the associated non-local variants of the  Y-system is postponed to Section~\ref{deltas}.
\begin{table}[tb]
\begin{center}
\begin{tabular}{|c|cc|}
\hline
Function & Singularity position &\\
\hline
\hline
$Y^{(\alpha)}_y(u)$  & $u=\pm 2 + i\frac{2J}{g}$, & $J= 0, \pm 1, \pm 2, \dots$\\
\hline
$ Y^{(\alpha)}_{(w|M)}(u)$ & & \\
\cline{1-1}
$Y^{(\alpha)}_{(v|M)}(u)$ & $u=\pm 2 + i\frac{J}{g}$, & $J= \pm M, \pm (M+2), \pm (M+4),\dots$ \\
\cline{1-1}
$ Y_{M}(u)$ & & \\
\hline
\end{tabular}
\caption{\small Square-root branch points for the  Y functions.}
\label{table1}
\end{center}
\end{table}
%
%%%%%%%%%%%%%%%%%%%%%%%%%%%%%%%%%%%%%%%%%%%%%%%%%%%%%%%%%%%%%%%%%%
%
\resection{TBA $\equiv$ dispersion relation}
\label{sectionDis}
{}From the works~\cite{Bombardelli:2009ns, Gromov:2009bc, Arutyunov:2009ur} it has emerged  that, contrary to the more studied relativistic invariant cases, the transformation from Y-system to TBA equations is not straightforward.
The  local form of the    $\Ad$-related  Y-system does not contain information on the branch points and in particular it is
almost totally insensitive to the precise form of  dressing factor. In this, and in the following  sections we shall  show that there exists a set of local functional constraints for the discontinuities that can
be directly  transformed  to   integral form. To explain   the idea
let us  first consider equation (\ref{eq:disctilde}). By setting
\eq
T(z,u) = \frac{1}{\sqrt{4-z^2}(z-u)} = {2\pi i \over  \sqrt{4-u^2}} K(z, u),
\label{T1}
\en
where
\eq
K(z-i \CQ/g, u) -K(z+i \CQ/g, u) = \phi_{\CQ,(y|-)}(z,u)- \phi_{\CQ,(y|+)}(z, u),
\label{K1}
\en
equation  (\ref{eq:disctilde})  can be written in the form
\eq
 {\ln \left( Y^{(\alpha)}_{(y|-)}(u)/Y^{(\alpha)}_{(y|+)}(u)\right) \over  \sqrt{4-u^2} }
=-\sum_{\CQ}\int_{\RR} {dz \over 2 \pi i} \, L_{\CQ}(z)( T(u-i\CQ/g, z)-T(u+i\CQ/g, z)  ).
\label{p1}
\en
We shall now   prove that  the relation on the discontinuities
\eq
\left[\ln \left(Y^{(\alpha)}_{(y|-)} /Y^{(\alpha)}_{(y|+)} \right) \right]_{ \pm 2N}=- \sum_{\CQ=1}^N [L_{\CQ}]_{\pm (2N-\CQ)},
\label{dis1}
\en
combined  with some more general  analyticity information, is equivalent to (\ref{p1}).
In  (\ref{dis1}),  $\ln$ denotes the principal branch of the complex logarithm   while  the  symbol $[f]_Z$ with  $Z \in \ZZ$  stands for the discontinuity of  $f(z)$ on the  semi-infinite segments described by $z=u+i Z/g$ with $u \in (-\infty,-2) \cup (2,+\infty)$:
\eq\label{eq:squarebrackets}
\left[ f \right]_{Z} =  \lim_{\ep \rightarrow 0^+} f(u+i Z/g+i \ep)- f(u+iZ/g -i \ep).
\en
Thus, the function
\eq
\left[ f \right(u)]_{Z}=f(u+i Z/g)- f(u_*+iZ/g)
\en
is the analytic extension of the discontinuity  (\ref{eq:squarebrackets}) to generic complex values of $u$.

We conjecture that  the relevance of  equation (\ref{dis1}) and the other   discontinuity relations  introduced   in the following
sections is not restricted to the   ground state  but, provided the analytic properties on the relevant  reference sheets are suitably  modified, they can be directly transformed
into   excited state TBA equations.

\begin{figure}[h]
\centering
\includegraphics[width=7cm]{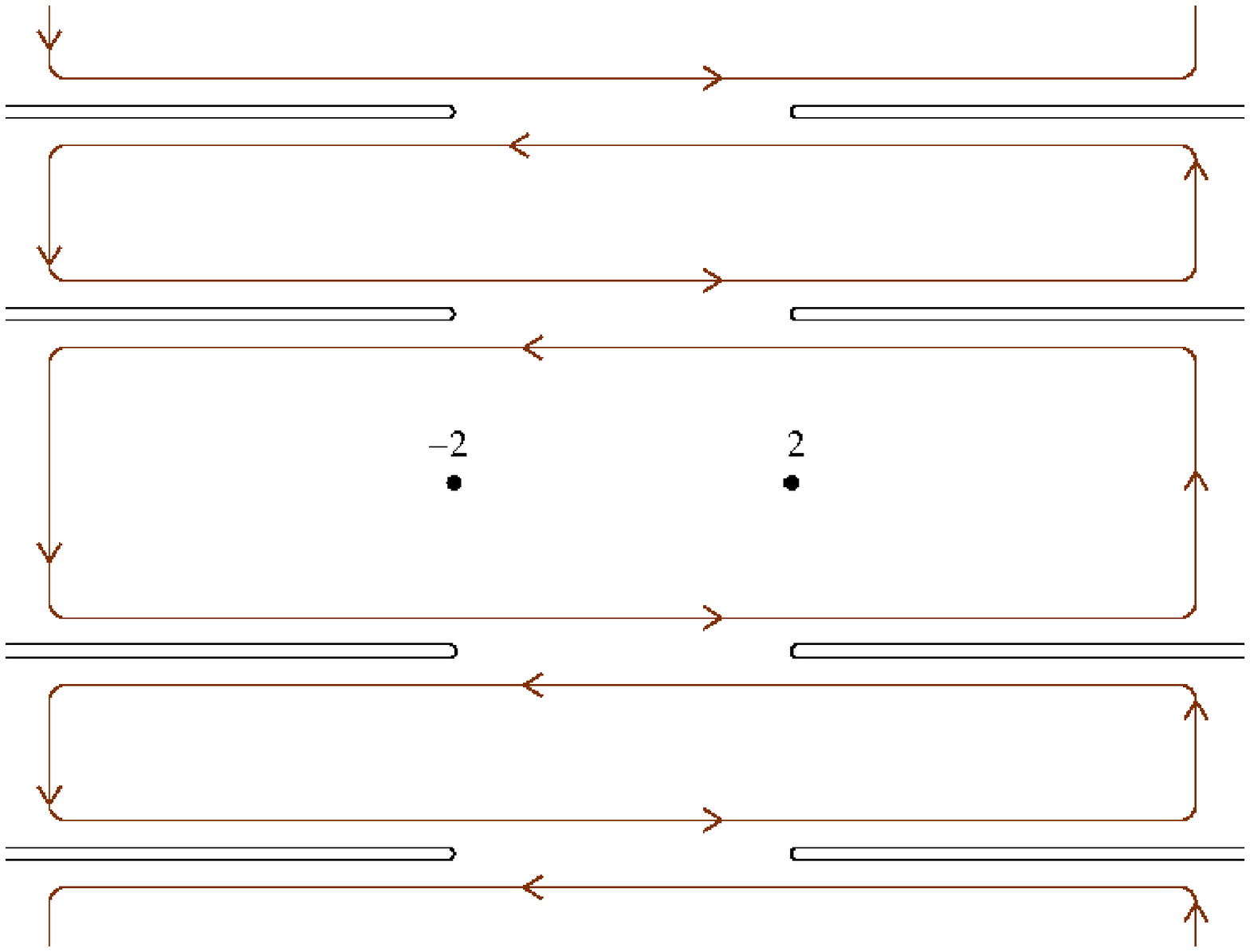}
\caption{ The deformed contour $\GammaO$. }
\label{Gamma0}
\end{figure}

First notice that     the quantity
appearing on the lhs of   (\ref{p1}) is analytic at the points $u=\pm 2$, but it still has an infinite set of  branch points at $u= \pm 2 \pm  i2 N/g$ with $N \in \mathbb{N}$. In the following we shall assume that, for the ground state
TBA equations,  these are the only singularities  of (\ref{p1})  on the first Riemann sheet.
By applying Cauchy's integral theorem we can first write
\eq
{\ln \left( Y^{(\alpha)}_{(y|-)}(u)/Y^{(\alpha)}_{(y|+)}(u)\right) \over  \sqrt{4-u^2} }
= \oint_{\gamma}  {dz \over 2 \pi i}  \frac{\ln\left( Y^{(\alpha)}_{(y|-)}(z)/Y^{(\alpha)}_{(y|+)}(z)\right)  }{(z-u)\sqrt{4-z^2} },
\label{p3}
\en
where $\gamma$ is a positive oriented contour running inside  the strip $ |\IIm(u)|< 1/g$, and then deform
$\gamma$ into the homotopically equivalent contour  $\GammaO$ represented in Figure~\ref{Gamma0} as the union of an infinite number of rectangular contours  lying between    the branch cuts of (\ref{p1}).
Since  $ \ln(Y^{(\alpha)}_{(y|-)}/Y^{(\alpha)}_{(y|+)})  \rightarrow O(1)$ uniformly as $|u| \rightarrow \infty$ the  sum of the  vertical segment contributions vanishes as the horizontal size of the rectangles tends to infinity.  Using  relation (\ref{dis1}) we can write the rhs of (\ref{p3})  as
\eq
 -\sum_{J,\s} \left(\int_{\RR -i2{\s}J/g +i \ep } - \int_{\RR -i2{\s}J/g -i \ep } \right) \frac{dz}{2 \pi i}
\sum_{\CQ=1}^J \frac{L_{\CQ}(z+i{\s}\CQ/g)  }{\sqrt{4 -
z^2} (z - u)},
\label{eq:cauchytilde}
\en
$(\s=\pm 1, J=1,2,\dots)$. However, as a consequence of the  identity
\eq
\left(\int_{\mathbb{R}+i\epsilon} -\int_{\RR - i\epsilon} \right)  dz \;\left( f(z
+\frac{i2}{g})+f(z)  \right)=\left( \int_{\RR +
\frac{i2}{g}+ i\epsilon} -\int_{\mathbb{R}-i\epsilon} \right) dz \; f(z),
\label{prop}
\en
which holds  for any function  $f(u)$ analytic in the strip $0<\IIm(u)<\frac{2}{g}$ and vanishing  for
$|\RRe(u)| \rightarrow \infty$ in this region,
 several cancelations take place in (\ref{eq:cauchytilde}).
 The property (\ref{prop}) and a  change of   integration variables $z \rightarrow z- i \s \CQ/g$ in (\ref{eq:cauchytilde}) leads to:
\begin{align}
\label{eq:passage}
&{\ln \left( Y^{(\alpha)}_{(y|-)}(u)/Y^{(\alpha)}_{(y|+)}(u)\right) \over  \sqrt{4-u^2} }
= \sum_{\CQ,\s}  \s \int_{\mathbb{R}+  i\s(\CQ/g - \epsilon)}
\frac{dz}{2 \pi i} \frac{ L_\CQ(z)}{\sqrt{4 - (z +i\s\CQ/g)^2} (z +
i\s\CQ/g - u)}\\
&=-\sum_{\CQ} \int_{\mathbb{R}} \frac{dz}{2 \pi i}
\left(\frac{ L_\CQ(z)}{\sqrt{4 - (z - i \CQ/g)^2}(z - i \CQ/g - u)}-\frac{ L_\CQ(z)}{ \sqrt{4 - (z +
i \CQ/g)^2} (z + i \CQ/g - u)}\right),\nn
\end{align}
which matches  the result (\ref{eq:disctilde}). To get the TBA equation (\ref{TBA2}) we still need to prove that
\eq
\ln\left({Y_{(y|-)}^{(\alpha)}(u)\text{ } Y_{(y|+)}^{(\alpha)}(u)} \right)
=\sum_{M} (2L^{(\alpha)}_{(v|M)}-2L^{(\alpha)}_{(w|M)}-L_{M})*\phi_M(u).
\label{eq:halfTBA}
\en
Let us show that (\ref{eq:halfTBA}) can be obtained as a dispersion relation as well. We simply use the fact (see Appendix~\ref{AE}) that the functional constraints (\ref{dis1}), together with the Y-system, imply:
\eq
\left[\ln\left({Y_{(y|-)}^{(\alpha)}\text{ } Y_{(y|+)}^{(\alpha)}} \right)\right]_{\pm 2N}=\sum_{J=1}^{N}\left[2L^{(\alpha)}_{(v|J)}-2L^{(\alpha)}_{(w|J)}-L_J\right]_{\pm(2N-J)}.
\en
Since $\ln\left({Y_{(y|-)}^{(\alpha)}Y_{(y|+)}^{(\alpha)}}\right)$ is regular in the strip $|\IIm(u)|<2/g$, by Cauchy's formula we finally arrive to  the desired result (\ref{eq:halfTBA}):
\begin{align}
\ln&\left({Y_{(y|-)}^{(\alpha)}(u)\; Y_{(y|+)}^{(\alpha)}}(u)\right)=-\sum_{M,\s}  \s \int_{\mathbb{R}+  i\s(M/g - \epsilon)} \frac{dz}{2 \pi i}
\frac{2L^{(\alpha)}_{(v|M)}(z)-2L^{(\alpha)}_{(w|M)}(z)-L_{M}(z)}{(z+i\s M/g - u)}\\
&=\sum_{M} \int_{\mathbb{R}}\frac{dz}{2 \pi i} {\Big (} 2L^{(\alpha)}_{(v|M)}(z)-2L^{(\alpha)}_{(w|M)}(z)-L_{M}(z){ \Big )}\left(\frac{1}{(z - i M/g - u)}-\frac{1}{(z + i M/g - u)}\right). \nn
\end{align}
All the kernels appearing in the system of TBA equations are
basically certain linear combinations of $K(z,u)$ and $\phi_M(u)$. Therefore, the derivation of
the rest of the TBA system from functional equations for discontinuities goes under
the same spell.
%
%%%%%%%%%%%%%%%%%%%%%%%%%%%%%%%%%%%%%%%%%%%%%%%%%%%%%%%%%%%%%%%%%%%%%%%%%%%%%%%%%%
%
\resection{The discontinuity function $\Delta(u)$}
\label{deltas}
Following closely the method described in  the previous section, we shall now derive  a   spectral representation for the function
\eq
\Delta(u) =\left[\ln Y_1(u)\right]_{+1}  = \ln \frac{Y_{(1|-)}(u+\frac{i}{g})}{Y_{(1|+)}(u+\frac{i}{g})},
\label{delta}
\en
where we have set $Y_{(1|-)}(u+\frac{i}{g})\equiv Y_{1}(u+\frac{i}{g})$, $Y_{(1|+)}(u+\frac{i}{g}) \equiv Y_{1}(u_*+\frac{i}{g})$, and the points $u$ and $u_*$ are analytically connected through the path depicted in Figure~\ref{cutfinal}. In other words, this function encodes the branching properties of $Y_1$ around $-2+i/g$, and can be related to the analytic behaviour around $-2 -i/g$ through the Y-system relation (\ref{eq:YQf}). Using the fact that
$Y_1(u-\fract{i}{g})Y_1(u+\fract{i}{g}){\prod_{\alpha}\left(1+\frac{1}{Y_{(y|-)}^{(\alpha)}(u)}\right)}=\left(1+Y_{2}(u)\right)$ is regular on the real axis, we find:
\eq
\bar{\Delta}(u) =\left[\ln Y_1(u)\right]_{-1}  = \ln \frac{Y_{1}(u-\frac{i}{g})}{Y_{1}(u_*-\frac{i}{g})}=
-\Delta(u)-\sum_{\alpha}\ln\frac{\left(1+\frac{1}{Y_{(y|-)}^{(\alpha)}(u)}\right)}{\left(1+\frac{1}{Y_{(y|+)}^{(\alpha)}(u)}\right)}.
\label{bdelta}
\en
It is interesting to recall how the function $\Delta$ was first introduced in~\cite{Arutyunov:2009ur}. A non local variant of the Y-system equation (\ref{eq:YQf}) is obtained if the function $Y_1(u)$ is shifted by $\pm i/g$ starting from a point in the region $|\RRe(u)|>2$, $|\IIm(u)|<1/g$. In this way, we are clearly forced to pass either of the branch cuts $u \in (-\infty, -2)\cup(2, \infty) \pm i/g$, so that, for $\IIm(u)>0$, on the lhs we get the product $Y_1(u-\fract{i}{g})Y_1(u_*+\fract{i}{g})=Y_{(1|-)}(u-\fract{i}{g})Y_{(1|+)}(u+\fract{i}{g})$. Using (\ref{eq:YQf}) and the definition {(\ref{delta}), we see that this function satisfies
\eq
Y_1(u-\fract{i}{g})Y_1(u_*+\fract{i}{g})={\left(1+Y_{2}(u)\right)\over {\prod_{\alpha}\left(1+\frac{1}{Y_{(y|-)}^{(\alpha)}(u)}\right)}}e^{-\Delta(u)},
\en
which is non local, as we will see in a moment. Reconnecting with the discussion of Section~\ref{Ysystem}, we stress that this relation cannot be obtained from (\ref{eq:YQf}) by following any analytic continuation path. The correct analytic continuation of (\ref{eq:YQf}) is rather
\eq
Y_1(u_*-\fract{i}{g})Y_1(u_*+\fract{i}{g})={\left(1+Y_{2}(u)\right)\over {\prod_{\alpha}\left(1+\frac{1}{Y_{(y|+)}^{(\alpha)}(u)}\right)}},
\label{diff}
\en
and, since $Y_1( (u+\frac{2 i}{g})_*-\frac{i}{g})\neq Y_1(u_*+\frac{i}{g})$, the functions on the lhs of (\ref{diff})  are now from two different sheets.

In~\cite{Arutyunov:2009ur} it was shown that, for $0<\IIm(u)<\frac{1}{g}$, $\Delta(u)$ admits  the following integral representation:
\begin{align}
\begin{split}
\Delta(u)=\sum_{N,\alpha} \int_{\mathbb{R}} dz &\; L^{(\alpha)}_{(v|N)}(z)(K(z+\frac{i}{g}N, u) + K(z - \frac{i}{g} N, u))\\
&+L \ln x^2(u)-\sum_{\alpha} L_{(y|-)}^{(\alpha)}(u)+\sum_{\alpha}  L_y^{(\alpha)}\so \;K( u)+
\Delta^{\Sigma}(u).
\end{split}
\label{eq:deltatba}
\end{align}
Equation (\ref{eq:deltatba}) needs to be modified when $u$ is moved outside the region $0<\IIm(u)<\frac{1}{g}$. For example, one has to use the property
\eq
L_{y}^{(\alpha)}\so \; K( u) \rightarrow -  L_{y}^{(\alpha)}\so \; K( u) + L_{(y|+)}^{(\alpha)}(u) + L_{(y|-)}^{(\alpha)}(u)
\en
under the analytic continuation  $u \rightarrow u_*$ to check that  $\Delta(u_*)=-\Delta(u)$, as expected from the definition (\ref{delta}).
Finally, in (\ref{eq:deltatba}) we have defined
\eq
\Delta^{\Sigma}(u) = 2\sum_{\CQ} L_{\CQ}\ast K_{\CQ}^{\Sigma}(u),
\en
and
\eq
K_{\CQ}^{\Sigma}(z,u) = K_{\CQ,1}^{\Sigma}(z,u+i/g)-K_{\CQ,1}^{\Sigma}(z,u_* +i/g).
\en
An explicit expression for the kernel\footnote{To relate this kernel to the following definition given in~\cite{Arutyunov:2009ur}:
\[
\check{K}^{\Sigma}_{\CQ}(z, u)=K_{\CQ,1}^{\Sigma}(z,u_*+i/g)+K_{\CQ,1}^{\Sigma}(z,u-i/g)-K_{\CQ,2}^{\Sigma}(z, u),
\]
it is sufficient to use property (\ref{DF}), yielding: $K_{\CQ}^{\Sigma}(z,u)=-\check{K}_{\CQ}^{\Sigma}(z,u)$.
} $K_{\CQ}^{\Sigma}$  requires the knowledge of the dressing factor in the mirror kinematics, which was found in~\cite{Arutyunov:dressingfactor}. Stemming from the result contained in \cite{Arutyunov:simplified}, in Appendix~\ref{AB} we show that $\Delta^{\Sigma}$ admits the following representation (see equation (\ref{eq:deltasigma3})):
\eq
\Delta^{\Sigma}(u)=\sum_{M,\alpha} \oint_{\bgammax} dz \;\ln Y_y^{(\alpha)}(z)\left(-K(z+i2M/g, u)+K(z-i2M/g, u)\right).
\en
The function $\Delta(u)$ has a branch point at every $u=\pm2 +i2Z/g$, $Z \in \mathbb{Z}$. The discontinuities across the branch cuts lying away from the real axis fulfil the following  simple functional equations:
\eq
\left[\Delta(z)\right]_{\pm 2N}=\mp\sum_{\alpha}\left[ L_{(y|\mp)}^{(\alpha)}(z)+\sum_{M=1}^{N} L^{(\alpha)}_{(v|M)}\left(z \mp \frac{iM}{g}\right)\right]_{\pm 2N}\mp\sum_{\alpha}\left[\ln{Y_{y}^{(\alpha)}(z)}\right]_{0},
\label{eq:discontinuities}
\en
($N=1,2, \dots$). Moreover, assuming that the asymptotic behaviour is dominated by the energy term in (\ref{eq:deltatba}), we see that:
\begin{align}
\begin{split}
e^{\Delta(u)}&\sim u^{2L} \rightarrow \infty~~~~~(\text{for}~ u\rightarrow \infty,~~\IIm(u)<0);\\
e^{\Delta(u)}&\sim 1/u^{2L} \rightarrow 0~~~~(\text{for}~ u\rightarrow \infty,~~\IIm(u)>0),
\end{split}
\end{align}
and an extra  branch cut appears  when taking the logarithm. Notice that, in order for (\ref{eq:discontinuities}) to be true, we are forced to draw this cut in the interior of the first sheet: we will assume that $\Delta(u)$ has a constant discontinuity running along the imaginary axis:
\eq
\Delta(iv+\epsilon)-\Delta(iv-\epsilon)=i2 L  \pi,~ (v\in \RR ).
\label{eq:constantdisc}
\en
(This corresponds to the principal branch prescription for $\ln x^2$ in (\ref{eq:deltatba}).)
Let us now show that the local information (\ref{eq:discontinuities}) and (\ref{eq:constantdisc}) is enough to recover equation (\ref{eq:deltatba}). Again, we start from Cauchy's integral formula for $\Delta(u)/\sqrt{4-u^2}$, with $u$ lying in the region $|\IIm(u)|<\frac{1}{g}$. A  positive oriented path $\gamma$ encircling $u$
can be  replaced by the contour $\GammaX$ represented in Figure~\ref{Gamma1} plus  two  vertical  lines on both sides of the logarithmic   cut~(\ref{eq:constantdisc}):
\begin{align}
\begin{split}
\frac{\Delta(u)}{\sqrt{4-u^2}}&=\oint_{\GammaX} \frac{dz}{2 \pi i}  \frac{\Delta(z)}{(z -u)\sqrt{4-z^2}}+\left(\int_{i\mathbb{R}-\epsilon}-\int_{i\mathbb{R}+\epsilon}\right) \frac{dz}{2 \pi i} \frac{\Delta(z)}{(z -u)\sqrt{4-z^2}} \\
&=I_{\GammaX}(u)-\frac{L}{\sqrt{4-u^2}}\int_{i\mathbb{R}} dz \; \frac{d}{dz}\ln\left(\frac{x(z)-x(u)}{x(z)-\frac{1}{x(u)}}\right)=I_{\GammaX}(u)
+\frac{L \ln{x^2(u)}}{\sqrt{4-u^2}},~~
\end{split}
\label{eq:cauchydelta}
\end{align}
and we see that, not only equation (\ref{eq:constantdisc})  signals  that the  quantisation of  $L$ is deeply related to
reflection symmetry of   $e^{\Delta(u)}$ about the imaginary axis but it also naturally  leads
to the  term  $L \ln x^2(u)$  in (\ref{eq:deltatba}) which is directly related to the  `driving' terms  $L\tilde{E}_{\CQ}(u)$ appearing in the TBA equations.
In the same context, but from a slightly different perspective, the quantisation of $L$   has been also discussed in~\cite{Frolov:2009in}.
\begin{figure}[h]
\centering
\includegraphics[width=7cm]{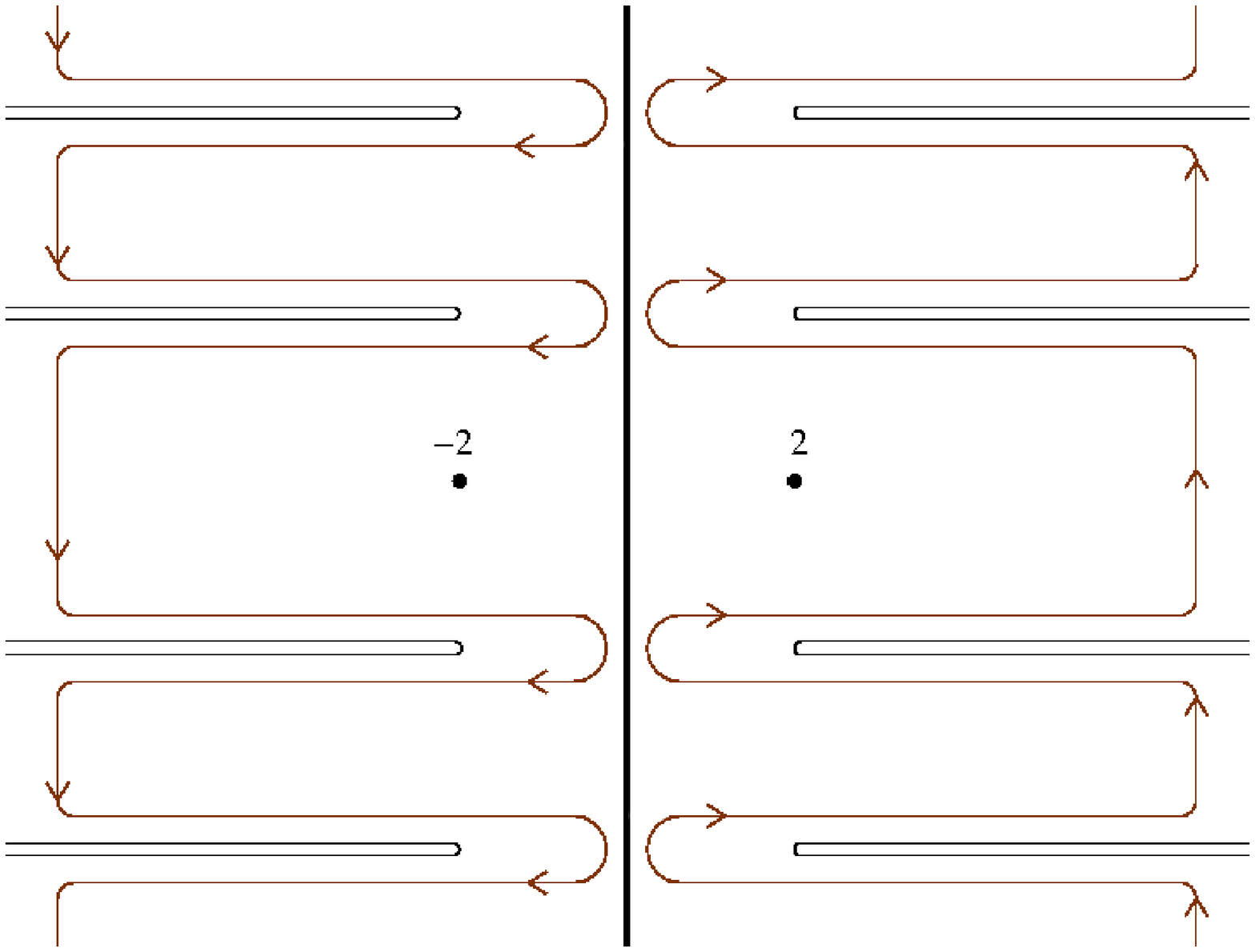}
\caption{ The  contour $\GammaX$: the vertical thick line corresponds to  the logarithmic cut of  $\Delta(u)$.  }
\label{Gamma1}
\end{figure}
Next,   using (\ref{eq:discontinuities}), the integral  $I_{\GammaX}(u)$ can be rewritten as:
\eq
I_{\GammaX}(u)=-\sum_{N,\alpha,\s} \s\oint_{\bgammax} \frac{dz}{2 \pi i} \frac{
 L_{(y|-\s)}^{(\alpha)}(z+i2\s N/g)+\sum_{M=1}^{N}L^{(\alpha)}_{(v|M)}(z+i\s(2N-M)/g)+\ln{Y_{y}^{(\alpha)}(z)}}{(z+i2\s N/g -u)\sqrt{4-(z+i2\s N/g)^2}}.
\label{quan}
\en
For $0<\IIm(u)<\frac{1}{g}$,
demanding  that $L^{(\alpha)}_{(v|M)}(u)$ is  regular on the whole first sheet and that $L_{(y|-)}^{(\alpha)}(u)$  and $L_{(y|+)}^{(\alpha)}(u)$ are regular respectively for   $\IIm(u)>0$ and $\IIm(u)<0$ , we get:
\begin{align}
I_{\GammaX}(u)=&\frac{1}{\sqrt{4-u^2}}\sum_{\alpha}{\Big (}\int_{\RR+i(2/g-\epsilon)} dz \; L^{(\alpha)}_{(y|-)}(z)K(z, u)+\int_{\RR-i(2/g-\epsilon)} dz \; L^{(\alpha)}_{(y|+)}(z)K(z, u)\nn\\
&+\sum_{N} \int_{\RR} dz \;L^{(\alpha)}_{(v|N)}(z)\left(K(z +iN/g, u)+K(z -iN/g, u)\right) \label{eq:deltacalculation}\\
&+\int_{\bgammax} dz \; \ln{Y_{(y|-)}^{(\alpha)}(z)} \sum_{N} \left(-K(z +i2N/g, u)+K(z -i2N/g, u)\right){\Big )}. \nn
\end{align}
We see that the last two lines already reproduce the convolutions with $v$-related functions and the term $\Delta^{\Sigma}$ in (\ref{eq:deltatba}). To conclude the derivation we use the property  $K(z, u)=-K(z_*, u)$, and replace the two integrals over $\RR \pm i(2/g-\epsilon)$ in the first line of (\ref{eq:deltacalculation}) with a single  convolution along $\bgammao$, plus a residue due to the pole of $K(z, u)$ at $z=u$ (see, Figure~\ref{Def}):
\begin{align}
\int_{\mathbb{R}+i(2/g-\epsilon)} dz \; L^{(\alpha)}_{(y|-)}(z)K(z, u) +\int_{\mathbb{R}-i(2/g-\epsilon)} dz \; L^{(\alpha)}_{(y|+)}(z)K(z, u)&   \label{res} \\
=\int_{\mathbb{R}+i(2/g-\epsilon)} dz \; L^{(\alpha)}_{y}(z)K(z, u)
-\int_{\mathbb{R}-i(2/g-\epsilon)} dz \; L^{(\alpha)}_{y}(z_*)K(z_*, u)&=- L_{(y|-)}^{(\alpha)}(u)+  L_y^{(\alpha)}\so \; K(u). \nn
\end{align}
Using (\ref{res}) and (\ref{eq:deltacalculation}) in (\ref{eq:cauchydelta}) we finally arrive to  (\ref{eq:deltatba}).
\begin{figure}[h]
\centering
\includegraphics[width=12.cm]{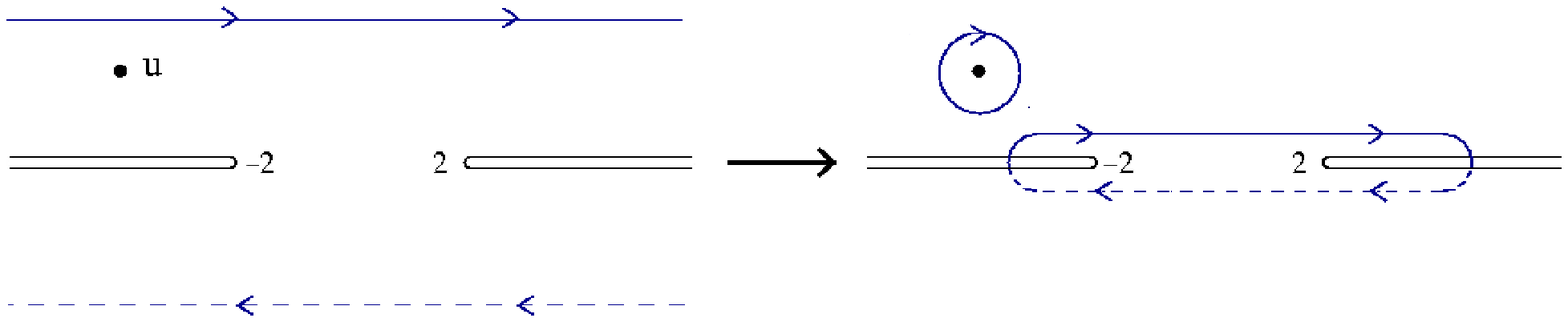}
\caption{ The contour deformation corresponding to equation (\ref{res}). }
\label{Def}
\end{figure}
%
%%%%%%%%%%%%%%%%%%%%%%%%%%%%%%%%%%%%%%%55
%
\resection{$w$-particles}
\label{wnodes}
As revealed by the TBA equation (\ref{TBA4}), the $Y^{(\alpha)}_{(w|M)}$-functions are free of singularities in the strip $|\IIm(u)|< M/g$, with square root branch points at the edge of this region. The discontinuities of $\ln Y_{(w|1)}^{(\alpha)}$ take the very simple form:
\eq
\left[\ln{Y^{(\alpha)}_{(w|1)}} \right]_{\pm 1}= \left[ L_{y}^{(\alpha)} \right]_0.
\label{eq:firstw}
\en
When supplemented with this simple information, the Y-system equation (\ref{yw}) is sufficient to reconstruct the whole set of discontinuities of the $w$-related  functions, and thus to define a dispersion relation. Taking the logarithm of the Y-system equation we have:
\eq
\ln Y_{(w|M)}^{(\alpha)}(u) =\sum_N \I_{M N} \La_{(w|N)}^{(\alpha)}(u-\fract{i}{g})+\delta_{M, 1} \left (L^{(\alpha)}_{(y|-)}(u -\fract{i}{g})-L^{(\alpha)}_{(y|+)}(u-\fract{i}{g}) \right )-\ln{Y_{(w|M)}^{(\alpha)}(u-\fract{i2}{g})}.
\label{eq:logyw}
\en
First, we notice that from the above stated analytic properties we can deduce that $\ln Y_{(w|M)}^{(\alpha)}(u)$ is regular except possibly at the points $u=-2\pm i(M+2I)/g$, $u=2\pm i(M+2I)/g$, with $I=0, 1, \dots$. At these values of $u$, (\ref{eq:logyw}) imposes a constraint on the monodromy properties of $\ln Y_{(w|M)}^{(\alpha)}$. A very useful relation is obtained after $M+I$ iterations of  (\ref{eq:logyw}):
\begin{align}
\begin{split}
\ln Y^{(\alpha)}_{(w|M)}(u \pm &i(M+2I)/g)  \\
=&{\Big ( } L^{(\alpha)}_{(y|-)}(u\pm i2I/g)-L^{(\alpha)}_{(y|+)}(u\pm i2I/g)+\sum_{N=1}^{M-1}  L^{(\alpha)}_{(w|N)}(u\pm i(2I+N)/g){\Big )}\\
&+\sum_{J=1}^{I}{\Big ( } L^{(\alpha)}_{(w|J)}(u\pm i(2I-J)/g)+\sum_{N=1}^{M-1} 2 L^{(\alpha)}_{(w|J+N)}(u\pm i(2I-J+N)/g)\\
&+L^{(\alpha)}_{(w|M+J)}(u\pm i(M+2I-J)/g){\Big )}+ \La^{(\alpha)}_{(w|M+I+1)}(u\pm i(M+I-1)/g)  \\
&+\sum_{N=1}^{M-1} L^{(\alpha)}_{(w|N+I+1)}(u \pm i(I+N-1)/g)-\ln Y^{(\alpha)}_{(w|I+1)}(u \pm i(I-1)/g).
\end{split}
\label{eq:iterateY}
\end{align}
To simplify the notation, let us rewrite (\ref{eq:iterateY}) through the definition of a family of functions $D^{(w|M)}_{\pm (M+2I)}$ \footnote{Here and in Section \ref{vnodes}  the label ``$\alpha$" is omitted for the sake of legibility: $D^{(w|M)}_{\pm (M+2I)} \equiv D^{(w|M),(\alpha)}_{\pm (M+2I)}$, $D^{(v|M)}_{\pm (M+2I)} \equiv D^{(v|M),(\alpha)}_{\pm (M+2I)}$.}:
\begin{align}
\begin{split}
\ln Y^{(\alpha)}_{(w|M)}(u &\pm i(M+2I)/g)=D^{(w|M)}_{\pm (M+2I)}(u)+\La^{(\alpha)}_{(w|M+I+1)}(u\pm i(M+I-1)/g)\\
&+ \sum_{N=1}^{M-1} L^{(\alpha)}_{(w|N+I+1)}(u \pm i(I+N-1)/g)-\ln Y^{(\alpha)}_{(w|I+1)}(u \pm i(I-1)/g).
\end{split}
\label{eq:definitionD}
\end{align}
By using (\ref{eq:firstw}) and the fact that {$\ln Y_{(w|N)}^{(\alpha)}(u)$} is regular for  $|\IIm(u)|<N/g$, we see that they are tied in the following way to the discontinuities of $w$-related functions:
\begin{align}
\Big[ \ln Y_{(w|M)}^{(\alpha)}(u &\pm  i(M+2I)/g) \Big]_{0} =\left[D^{(w|M)}_{\pm(M +2I)}(u)\right]_{0} -{\Big [} \ln Y^{(\alpha)}_{(w|I+1)}(u \pm i(I-1)/g){ \Big ]}_0\nn\\
&+{\Big [} \La^{(\alpha)}_{(w|M+I+1)}(u\pm i(M+I-1)/g)+\sum_{N=1}^{M-1} L^{(\alpha)}_{(w|N+I+1)}(u \pm i(I+N-1)/g){\Big ]}_0 \nn\\
=&\left[D^{(w|M)}_{\pm(M +2I)}(u)\right]_{0}-\delta_{I, 0}{\Big [}\ln Y^{(\alpha)}_{(w|1)}(u \mp i/g){ \Big ]}_0 \label{eq:Dw} \\
=&\left[D^{(w|M)}_{\pm(M +2I)}(u)\right]_{0}-\delta_{I, 0}{\Big [} L^{(\alpha)}_{y}(u){ \Big ]}_0. \nn
\end{align}
The comparison between (\ref{eq:iterateY}) and (\ref{eq:definitionD}) gives an explicit expression for these functions. For example:
\eq
D^{(w|M)}_{\pm M}(u)= L_{(y|-)}^{(\alpha)}(u)-L_{(y|+)}^{(\alpha)}(u) +\sum_{N=1}^{M-1} L_{(w|N)}^{(\alpha)}(u \pm iN/g).
\label{eq:firstwfunction}
\en
All the other functions in the family can be obtained using the following rule, which is also very significant in view of property (\ref{prop}):
\begin{align}
\begin{split}
D^{(w|M)}_{\pm(2J+M)}(u)-D^{(w|M)}_{\pm(2J+M-2)}(u\pm i2/g)=& 2\sum_{N=J+1}^{M+J-1} L_{(w|N)}^{(\alpha)}(u \pm iN/g)+L^{(\alpha)}_{(w|J)}(u\pm iJ/g) \\
&+ L^{(\alpha)}_{(w|M+J)}(u\pm i(M+J)/g),
\end{split}
\label{eq:rulew}
\end{align}
($J=1,2,\dots$).
Let us show how to derive the TBA equation from this information. From (\ref{eq:Dw}) and (\ref{eq:firstwfunction}) we see that, with $|\IIm(u)|< M/g$ on the first Riemann sheet, Cauchy's formula reads:
\begin{align}
\ln Y_{(w|M)}^{(\alpha)}(u)=&
-\oint_{\bgammax}\frac{dz}{2\pi i}  \left(\frac{L_{y}^{(\alpha)}(z)}{(z-u+iM/g)}+\frac{L_{y}^{(\alpha)}(z)}{(z-u-iM/g)}\right)\nn\\
&+\sum_{I=0}^{\infty}\sum_{\s} \oint_{\bgammax} \frac{dz}{2\pi i} \frac{D^{(w|M)}_{\s(M+2I)}(z)}{(z-u+i\s(M+2I)/g)}\nn\\
=&-\oint_{\bgammax} \frac{dz}{2\pi i}  \left(\frac{L_{y}^{(\alpha)}(z)}{(z-u+iM/g)}+\frac{L_{y}^{(\alpha)}(z)}{(z-u-iM/g)}\right)\nn\\
&-{\Big(}\int_{\mathbb{R}-i\epsilon}\frac{dz}{2\pi i(z-u+iM/g)}-\int_{\mathbb{R}+i\epsilon}\frac{dz}{2\pi i(z-u-iM/g)}{\Big)}(L_{(y|-)}^{(\alpha)}(z)-L_{(y|+)}^{(\alpha)}(z)) \nn\\
&-\sum_{\s} \s\int_{\mathbb{R}-i \s\epsilon}\frac{dz}{2\pi i(z-u+i\s M/g)}\sum_{N=1}^{M-1}L_{(w|N)}^{(\alpha)}(z +i\s N/g)\\
&-\sum_{N=1}^{\infty}\sum_{\s} \s\int_{\mathbb{R}-i\s\epsilon} \frac{dz}{2\pi i} \frac{(D^{(w|M)}_{\s(M+2N)}(z)-D^{(w|M)}_{\s(M+2N-2)}(z+i2\s/g))}{(z-u+i\s(M+2N)/g)}.\nn
\end{align}
As anticipated, now we just need to insert (\ref{eq:rulew}) into the last line. Moreover, we can use the property (\ref{eq:property}) to reassemble all the convolutions with $y$-related functions into a single integral along $\bgammao$, and finally we find:
\begin{align}
\ln Y_{(w|M)}^{(\alpha)}(u)=&  L^{(\alpha)}_y \so \; \phi_M(u) +  \sum_{N=1}^{M-1} L_{(w|N)}^{(\alpha)}* \phi_{M-N}(u)-\sum_{J,\s} \s\int_{\mathbb{R}}\frac{dz}{2 \pi i (z-u+i\s(M+2J)/g)}\nn\\
&\times{\Big (}L^{(\alpha)}_{(w|J)}(z+i\s J/g)+L^{(\alpha)}_{(w|M+J)}(z+i\s(M+J)/g)+2\sum_{N=J+1}^{M+J-1} L_{(w|N)}^{(\alpha)}(z+ i\s N/g){\Big)} \nn\\
=&L^{(\alpha)}_y \so \;\phi_M(u) +\sum_{N} L^{(\alpha)}_{(w|N)}*\phi_{NM}(u). \label{int}
\end{align}
The result (\ref{int}) perfectly matches the TBA equation (\ref{TBA4}).
%
%%%%%%%%%%%%%%%%%%%%%%%%%%%%%%%%%%%%%%%%%%%%%%%%%%%%%%%%%%%%%%%%%%%%%%%
%
\resection{ $v$-particles}
\label{vnodes}
For $v$-related functions the situation is very similar to that encountered in the previous section. The essential information is encoded in the statement that $\ln Y^{(\alpha)}_{(v|M)}(u)$ is analytic for $|\IIm(u)|<M/g$ and in the value of the discontinuities closest to the real line:
\eq
\left[\ln{Y^{(\alpha)}_{(v|1)}} \right]_{\pm 1}= \left[ \La_{y}^{(\alpha)}\right]_0,
\label{eq:firstv}
\en
Also in this case, the Y-system leads to the introduction of a set of functions $D^{(v|M)}_{\pm(M+2I)}$, ($I=0, 1, \dots$), constrained by the following relations
\begin{align}
D^{(v|M)}_{\pm M}(u) = \La_{(y|-)}^{(\alpha)}(u)
-\La_{(y|+)}^{(\alpha)}(u)+&\sum_{N=1}^{M-1} L_{(v|N)}^{(\alpha)}(u \pm iN/g),\nn\\
D^{(v|M)}_{\pm(2J+M)}(u)-D^{(v|M)}_{\pm(2J+M-2)}(u\pm i2/g) =&L^{(\alpha)}_{(v|J)}(u\pm iJ/g)+L^{(\alpha)}_{(v|M+J)}(u\pm i(M+J)/g) \label{eq:rulev} \\
&+2 \sum_{N=J+1}^{M+J-1}L_{(v|N)}^{(\alpha)}(u \pm iN/g)-\sum_{\CQ=J+1}^{M+J}L_{\CQ}(u \pm i\CQ/g),\nn
\end{align}
and satisfying
\eq
\left[ \ln Y^{(\alpha)}_{(v|M)}\right]_{\pm M}=\left[D^{(v|M)}_{\pm M}(u)-\La_{y}^{(\alpha)}\right]_{0},\hspace{0.5cm}\left[ \ln Y^{(\alpha)}_{(v|M)}\right]_{\pm(M+2J)}= \left[D^{(v|M)}_{\pm(M +2J)}\right]_{0},
\label{eq:Dv}
\en
for ($J=1, 2, \dots$). Finally, using Cauchy's  theorem:
\begin{align}
\ln Y_{(v|M)}^{(\alpha)}(u)
=&\La^{(\alpha)}_y \so \; \phi_M(u)+\sum_{N} L^{(\alpha)}_{(w|N)}*\phi_{NM}(u) \nn\\
&+\sum_{J,\s} \s\int_{\mathbb{R}-i\s \epsilon}\frac{dz}{2 \pi i (z-u+i\s(M+2J))}\sum_{\CQ=J+1}^{M+J}L_{\CQ}(z +i\s\CQ/g)\nn\\
=&  L^{(\alpha)}_y \so \; \phi_M(u) +\sum_{N} L^{(\alpha)}_{(w|N)}*\phi_{NM}(u) \label{eqeq} \\
&+\ln Y_y^{(\alpha)} \so\;\phi_{M}(u) -\sum_{\CQ} \sum_{I=0}^{M-1} L_{\CQ}* \phi_{\CQ-M+2I}(u).\nn
\end{align}
Thanks to the identity (\ref{eq:vintermediate}), equation (\ref{eqeq})  is equivalent to the TBA equation (\ref{TBA3}).
%
%%%%%%%%%%%%%%%%%%%%%%%%%%%%%%%%%%%%%%%%%%%%%%%%%%%%%%%%%%%%%%%%%%%%%%%%%%%%%
%
\resection{$\CQ$-particles and the dressing kernel}
\label{Qnodes}
We already know from (\ref{delta},\ref{bdelta}) that the discontinuities of $\ln Y_1(u)$ at the border of the analyticity strip around the real axis can be linked via the following relations
\eq
\left[\ln Y_1(u)\right]_{+1}=\Delta(u)=\frac{1}{2}\left[\Delta(u)\right]_0,~~\left[\ln Y_1(u)\right]_{-1}=\bar{\Delta}(u)=-\frac{1}{2}\left[\Delta(u)\right]_0-\sum_{\alpha}\left[L_y^{(\alpha)}(u)\right]_0,
\label{Du}
\en
to the function $\Delta(u)$, defined through the integral representation (\ref{eq:deltatba}). It is precisely the non locality of $\Delta$ (as well as of the difference $\ln Y_{(y|-)}^{(\alpha)}-\ln Y_{(y|+)}^{(\alpha)}$, see (\ref{eq:disctilde})) which makes the $\CQ-$ and $y$-related functions drastically different from the other species of excitations, forcing us to lean on an additional infinite set of functional relations which are not obtainable from the Y-system: (\ref{dis1}) and (\ref{eq:discontinuities}) respectively.

In Section~\ref{deltas} we have shown how to retrieve the form of $\Delta$ from the discontinuity relations (\ref{eq:discontinuities}). The rest of the derivation of the TBA equation (\ref{TBA1}) is based solely on the Y-system and goes along the same way as for $w-$ and $v-$nodes.

We define the family of $D^{\CQ}_{\CQ+2\CI}$ functions through the
following recursion relation:
\begin{align}
\label{eq:recurrenceQ}
\begin{split}
D^{\CQ}_{\pm(\CQ+2\CJ)}(u)&-D^{\CQ}_{\pm(\CQ+2\CJ-2)}(u\pm
i2/g)= L_{\CJ}(u\pm i\CJ/g)+
2\sum_{\CQ'=\CJ+1}^{\CQ+\CJ-1} L_{\CQ'}(u\pm i\CQ'/g)  \\
&+L_{\CQ+\CJ}(u\pm
i(\CQ+\CJ)/g)-\sum_{\alpha}\sum_{N=\CJ}^{\CQ+\CJ-1}
L^{(\alpha)}_{(v|N)}(u\pm iN/g),
\end{split}
\end{align}
($\CJ =1,2,\dots$), with initial condition:
\bea\label{eq:firstQ}
D^{\CQ}_{\pm \CQ}(u)=\sum_{\CQ'=1}^{\CQ-1}L_{\CQ'}(u\pm i\CQ'/g)
-\sum_{\alpha} \left(\sum_{N=1}^{\CQ-1}L^{(\alpha)}_{(v|N)}(u \pm
iN/g)+L^{(\alpha)}_{y}(u)\right).
\eea
These functions are related to the monodromy
properties of the  $(Y_{\CQ})'s$ through the following:
\begin{align}
\label{eq:DQ}
\begin{split}
&\left[ \ln Y_{\CQ}(u)\right]_{+\CQ}=\left[D^{\CQ}_{\CQ}(u)+\ln Y_1(u+{i \over g})+\sum_{\alpha}L^{(\alpha)}_{y}(u)\right]_{0}, \\
&\left[ \ln Y_{\CQ}(u)\right]_{-\CQ}=\left[D^{\CQ}_{-\CQ}(u)-\ln Y_1(u+{i \over g})\right]_{0},\\
&\left[ \ln Y_{\CQ}(u)\right]_{\pm(\CQ+2\CJ)}=\left[D^{\CQ}_{\pm(\CQ +2\CJ)}(u)\right]_{0}, \hspace{1cm}\text{   (}\CJ =1, 2, \dots\text{)}.
\end{split}
\end{align}
Proceeding in the usual way, we apply Cauchy's method  to $\ln Y_{\CQ}(u)$, with $|\IIm(u)|<\CQ/g$, and, as a consequence of
 (\ref{eq:recurrenceQ},\ref{eq:firstQ},\ref{eq:DQ}), we obtain:
\begin{align}
\ln{Y_{\CQ}}(u)
=&-\oint_{\bgammax} dz \; \ln Y_1(z+\frac{i}{g})\phi_{\CQ}(z-u)+\sum_{\CQ'} L_{\CQ'}*\phi_{\CQ' \CQ}(u) -\sum_{\alpha}\sum_{N=1}^{\CQ-1}  L_{(v|N)}^{(\alpha)}*\phi_{\CQ-N}(u)\nn\\
&+\sum_{\alpha}{\Big(}\oint_{\bgammax}\frac{dz}{2 \pi i(z-u+i\CQ/g)}L_{(y|-)}^{(\alpha)}(z)-\sum_{\s}\s\int_{\mathbb{R}+i\s\epsilon}\frac{dz}{2 \pi i(z-u-i\s\CQ/g)}L_{(y|-)}^{(\alpha)}(z){\Big )}\nn\\
&-\sum_{\CJ,\alpha,\s} \s\int_{\mathbb{R}-i\s\epsilon}\frac{dz}{2\pi i(z-u+i\s(\CQ+2\CJ)/g)}\sum_{N=\CJ}^{\CQ+\CJ-1} L^{(\alpha)}_{(v|N)}(z+i\s N/g).
\end{align}
In the end, reordering the different terms we get
\begin{align}
\begin{split}
\ln{Y_{\CQ}}(u)
=&-\oint_{\bgammax} dz \; \ln Y_1(z+\frac{i}{g}) \phi_{\CQ}(z-u)+\sum_{\CQ'} L_{\CQ'}*\phi_{\CQ' \CQ}(u)\\
&-\sum_{\alpha} \int_{\RR + i \epsilon} dz \; L_{(y|-)}^{(\alpha)}(z)\phi_{\CQ}(z-u)-\sum_{N,\alpha}  \sum_{J=1}^{\CQ} L_{(v|N)}^{(\alpha)}*\phi_{N-\CQ+2J}(u).
\end{split}
\label{eq:cauchyQ}
\end{align}
This is a very peculiar equation. Notice that the only visible kernels are difference-type, while the genuinely non-relativistic elements of (\ref{TBA1}) are hidden  inside the first convolution
\eq
\oint_{\bgammax}dz \; \ln Y_1(z+\frac{i}{g}) \phi_{\CQ}(z-u)= \frac{1}{2}\oint_{\bgammax}dz \; \Delta(z) \phi_{\CQ}(z-u),
\en
and we need to use the result (\ref{eq:deltatba}) to unravel them. For example, the energy term comes naturally to the surface through the following dispersion relation:
\eq
-\int_{\bgammax} dz \; \ln{x(z)} \phi_{\CQ}(z-u)=\ln\frac{x(u-\frac{i \CQ}{g})}{x(u+\frac{i \CQ}{g})}\equiv \widetilde{E}_{\CQ}(u).
\en
A number of very similar computations is reported in Appendix~\ref{AD}, culminating in (\ref{eq:Y1convolution}). This result shows that equation (\ref{eq:cauchyQ}) reduces to the original TBA equation (\ref{TBA1}) on the condition that the following identity holds for the dressing-related part:
\eq
-\frac{1}{2} \; \Delta^{\Sigma} \sx \;\phi_{\CQ}(u) = 2\sum_{\CQ'}L_{\CQ'} * K_{\CQ' \CQ}^{\Sigma}(u).
\label{eq:dressingrelated}
\en
In Appendix~\ref{AB} we evaluate the lhs of this equation. The result is:
\eq
 \sum_{\CQ'} L_{\CQ'} \ast K^{\Sigma}_{\CQ'\CQ}(u)= \sum_{\CQ'} L_{\CQ'} \ast \oint_{\bgammax} d\z \; \phi_{\CQ', y}(\z) \oint_{\bgammax} d\zi \; K_{\Gamma}^{\left[2\right]}(\z- \zi)\phi_{y, \CQ}(\zi, u).
\label{eq:hofman2maintext2}
\en
Further discussion and a direct proof of this relation is given in Appendix \ref{AC}. As discussed there, (\ref{eq:hofman2maintext2}) realizes a manifest symmetry between the direct and mirror theories. This is the object of the next section.
%
%%%%%%%%%%%%%%%%%%%%%%%%%%%%%%%%%%%%%%%%%%%%%%%%%%%%%%%%%%%%%%%%%%%%%%%%%
%
\resection{Application: the TBA equations for the direct theory}
\label{TBADD}
A possible question is whether the TBA equations for the direct theory can be obtained by considering the same set of
discontinuity relations but with complementary  prescriptions for the cuts:  running parallel to the real axis but inside the strip $|\RRe(u)|<2$.
The result is particularly simple: the TBA equations remain formally unchanged but for the replacement  $\bgammao \rightarrow \bgammax$:
\begin{align}
\vep_{\CQ}(u) =\mu_{\CQ}+&R\, E_{\CQ}(u)
-\sum_{\CQ} L_{\CQ'}*\phi^{\sigma}_{\CQ' \CQ}(u)
+ \sum_{\alpha}  \Big(\sum_{M} L^{(\alpha)}_{v|M}*\phi^{\ts}_{(v|M),\CQ}(u)
+  L^{(\alpha)}_{y} \sx \;\phi_{y,\CQ}^{\ts}(u) \Big)
,\label{TBAS1}\\
\vep_{y}^{(\alpha)}(u)&=\mu_{y}^{(\alpha)}-\sum_{\CQ}  L_{\CQ}*\phi_{\CQ,y}^{\ts}(u)
+ \sum_{M} (L^{(\alpha)}_{(v|M)}-L^{(\alpha)}_{(w|M)}  )*\phi_{M}(u), \label{TBAS2}\\
\vep^{(\alpha)}_{(w|K)}(u)&=\mu_{(w|K)}^{(\alpha)}+\sum_{M} L^{(\alpha)}_{(w|M)}*\phi_{M K}(u)+
L^{(\alpha)}_{y} \sx \; \phi_{K}(u),
\label{TBAS4}\\
\vep^{(\alpha)}_{(v|K)}(u)&=\mu_{(v|K)}^{(\alpha)} -
\sum_{\CQ} L_{\CQ}*\phi_{\CQ,(v|K)}^{\ts}(u)
+\sum_{M} L^{(\alpha)}_{(v|M)}*\phi_{M K}(u) + L^{(\alpha)}_{y} \sx \;\phi_K(u)
,\label{TBAS3}
\end{align}
where $\alpha=1,2$, $K=1,2,\dots$,
\eq
\phi_{\CQ' \CQ}^{\sigma}(z,u) = -\phi_{\CQ' \CQ}(z-u) + 2 K^{\sigma}_{\CQ' \CQ}(z,u),~~K^{\sigma}_{\CQ'\CQ}(z,u)={1 \over 2 \pi i} \frac{d}{dz} \ln \sigma_{\CQ'\CQ}(z,u),
\label{dkd}
\en
and the new set of kernels $\{\phi_{y,\CQ}^{\ts},   \phi_{\CQ, y}^{\ts}, \phi_{(v|M),\CQ}^{\ts},
\phi_{\CQ, (v|M)}^{\ts}  \}$ is obtained from the kernels used in (\ref{TBA1}-\ref{TBA3})   by replacing the function $x$ with the
$x_{\ts}$  defined in (\ref{xsu}) and (\ref{xxs}). Finally, the energy and momentum are respectively
\eq
E_{\CQ}(u)= i g x_{\ts} \left(u- i \fract{\CQ}{g} \right)- ig x_{\ts} \left(u+ i \fract{\CQ}{g} \right)-\CQ,
\label{Edir}
\en
and
\eq
p^{\CQ} (u)=  i\ln \left ( {x_{\ts}(u-i \CQ/g) \over x_{\ts}(u+i \CQ/g) }  \right).
\en
An interesting observation is related to  the driving terms  (\ref{Edir}) appearing in the TBA equations. For the direct theory,
a   requirement analogous to (\ref{eq:constantdisc})
\eq
{\Delta(iv+\epsilon)-\Delta(iv-\epsilon) \over 2 \pi i } \in \ZZ,~~ (v\in \RR ),
\label{eq:constantdisc1}
\en
a part for the trivial case $\Delta(iv+\epsilon)=\Delta(iv-i\epsilon)$ cannot be straightforwardly  implemented due to the presence of an infinite number of  square-root branch cuts crossing the imaginary axis.  The result (\ref{Edir})
comes instead from the requirement
\eq
\frac{\Delta(u)}{\sqrt{4-u^2}} \sim R  + O(1/|u|)
\en
as $|u| \rightarrow \infty$.  The surprising consequence  of this fact is that, contrary to the mirror scale factor $L$,  there is no evident  mathematical advantage  to quantize $R$.
The steps that lead to (\ref{TBAS1}-\ref{TBAS3}) parallel completely the derivation of the TBA for the mirror theory starting
from the Y-system and the discontinuity relations. Therefore, automatically the solutions to (\ref{TBAS1}-\ref{TBAS3}) satisfy the
Y-system of Section~\ref{Ysystem}  which, in turn, can  easily be re-derived from (\ref{TBAS1}-\ref{TBAS3}) in the region $|\RRe(u)|>2$.
The latter simple observation motivates a further, perhaps more interesting, final comment. The TBA equations for the  direct $\Ad$ theory can be  directly derived from the Beisert and Staudacher equations~\cite{BS} through the string hypothesis and the TBA method. There are various variants of these Bethe Ansatz equations which differ  in the way  the momentum-related  factors $x_{\ts}(u-i\CQ/g)/x_{\ts}(u+i\CQ/g)$ are  rearranged   inside  the equations. From the result (\ref{dkd})
we  see that there in no trace of these  factors  in the kernels   $\phi_{\CQ' \CQ}^{\sigma}$. Therefore, contrary to our   initial expectations  equations (\ref{TBAS1}-\ref{TBAS3}) do not  descend, for  example,  from  equation  (6.4)  of~\cite{Arutyunov:2007tc} where the space length on which the TBA procedure  is based is the `{\bf R}-charge'  ${\it  J}$.  More precisely,
(6.4) of~\cite{Arutyunov:2007tc} leads to  TBA equations which differ from  (\ref{TBAS1}-\ref{TBAS3}) by  a scale dependent  set of chemical potentials  for the $\CQ$-particles
\eq
 \mu_{\CQ}(R) =  -\CQ \sum_{\CQ'}\int_{\RR} {du \over 2 \pi}  \,  {dp^{\CQ'} \over du}  L_{\CQ'}(u)=Q\tilde{E}_0(R),
\label{muq}
\en
where $\tilde{E}_0(R)$ is the ground state energy of the mirror theory.
Note that, since  the  chemical potentials (\ref{muq})   fulfil  the constraints (\ref{constmu}),  both versions of the TBA equations satisfy the  Y-system described in Section~\ref{Ysystem} and the discontinuity relations (\ref{d2}-\ref{d3}). They can be obtained by imposing different asymptotic conditions  on the pseudoenergies when  Cauchy's theorem is applied.
We suspect that the  form of  these additional R-dependent terms -if really needed-  should descend from  some independent  physical requirement, as for example, from the relation between the $\Ad$ Y-  and T- functions~\cite{Gromov:2009tv} and the transfer matrix of a possible underlying lattice system.
%%%%%%%%%%%%%%%%%%%%%%%%%%%%%%%%%%%%%%%%%%%%%
%
\resection{Conclusions}
\label{conclusions}
The analytic properties  of the solutions of the   $\Ad$ thermodynamic Bethe Ansatz equations are very different from those of the relativistic integrable quantum field theories. The Y functions  live on a complicated, and almost completely unexplored,   Riemann surface with an infinite number of square-root branch points.
The $\Ad$ Y-system   does not contain enough information to allow its   transformation into the integral form relevant for  $\Ad$ and in  particular  to reproduce the  dressing kernel. In this paper the analytic properties of the Y functions are studied in more detail
and the needed extra analytic information is identified  and encoded in a  new universal set of local functional
relations for the square-root discontinuities.
From these equations and the Y-system  the ground state   TBA equations for both the  mirror and the  direct theories were derived  using a novel method based on the Cauchy's  integral  theorem and the interpretation of the TBA equations as dispersion relations.

The  simplified set of TBA equations  obtained in~\cite{Arutyunov:2009ur,Arutyunov:simplified} can be easily derived from the extended Y-system and
Fourier transform   techniques can  be  used to short-cut further  some of the steps in the   Y$\rightarrow$ TBA inversion process.
Therefore, the apparently very-complicated structure of the $\Ad$
TBA equations is encoded into a fairly simple   set of local functional relations.

To get the TBA equations for the ground state we have adopted  certain minimality assumptions for the number of logarithmic
singularities in the reference Riemann sheet.
A main conjecture of this paper is that by the  same method,  including extra logarithmic singularities, one has  direct access to the full spectrum of the theory. Possibly, this will lead to  a  rigorous proof of the  TBA equations for the excited states conjectured in the
papers~\cite{Gromov:2009tv, Gromov:2009bc,  Arutyunov:2009ax}.  Work in this direction is  in progress.

Other sets of functional relations -the T-systems- play also a very fundamental   r\^ole in integrable models~\cite{Kirillov:1987zz,Kuniba:1992ev}.
In general Y- and T-systems are almost totally equivalent, but in  the $\Ad$-related cases~\cite{Gromov:2009tv},   T-systems
seem to contain more information compared to  the   basic Y-systems and they  can be considered more fundamental.
In this respect, it  would be important to  find  the analog of  equations~(\ref{d2}-\ref{d3}) for the discontinuities of the T-functions.

Our final  remark is that much  more work is needed to understand the integrable structure of the $\Ad$ thermodynamic Bethe Ansatz equations and in particular the Riemann surface associated to their solutions but we think that the results of this  paper provide
a further important  step in this direction.

\section*{Acknowledgements}
We would like to thank  Gleb Arutyunov, Zoltan Bajnok,  Sergey Frolov, Nikolay Gromov and  Vladimir Kazakov, for suggestions and very useful discussions.
We acknowledge the INFN grants IS PI14{\it ``Topics in non-perturbative gauge dynamics in field and string theory''} and PI11 for travel financial support, and the University PRIN 2007JHLPEZ ``Fisica Statistica dei Sistemi Fortemente Correlati all'Equilibrio e
Fuori Equilibrio: Risultati Esatti e Metodi di Teoria dei Campi".

%%%%%%%%%%%%%%%%%%%%%%%%%%%%%%%%%%%%%%%%%%%%%
%%%%%%%%%%%%%%%%%%%%%%%%%%%%%%%%%%%%%%%%%%%%%

\appendix
\resection{The S-matrix elements}
\label{AA}
Here we report  the scalar factors $S_{AB}$ involved in the definition of kernels in the TBA equations (\ref{TBA1}-\ref{TBA3}).
\eq
S_{y, \CQ}(u,z)=S_{\CQ, y}(z,u) =  \left(\frac{x(z-\frac{i}{g} \CQ)-x(u)}{x(z+\frac{i}{g}\CQ)-x(u)} \right) \sqrt{\frac{x(z+\frac{i}{g}\CQ)}{x(z-\frac{i}{g}\CQ)}}.
\label{SyQ}
\en
\begin{align}
\begin{split}
S_{(v|M), \CQ}(u,z)=&S_{\CQ,(v|M)}(z,u)= \left(
\frac{x(z-\frac{i}{g}\CQ)-x(u+\frac{i}{g}M )}{x(z+\frac{i}{g}\CQ)-x(u+ \frac{i}{g} M )}\right)
\left(\frac{x(z+\frac{i}{g}\CQ)}{x(z-\frac{i}{g}\CQ)} \right) \\
&\times \left(\frac{x(z-\frac{i}{g}\CQ)-x(u-\frac{i}{g} M)}{x(z+\frac{i}{g}\CQ)-x(u-\frac{i}{g}M )}\right)\prod_{j=1}^{M-1} \left(\frac{z-u-\frac{i}{g}(\CQ-M+2j)}{z-u+\frac{i}{g}(\CQ-M+2j)}\right),
\end{split}
\end{align}
\eq
S_M(u) =  \left(\frac{u-\frac{i}{g} M }{u+\frac{i}{g} M }
\right),
\en
\begin{align}
S_{K M}(u)&= \prod_{k=1}^{K}\prod_{l=1}^{M} S_{((K+2-2k)-(M-2l))}(u)  \\
&= \left( {u - \fract{i}{g} |K-M| \over u +\frac{i}{g} |K-M|} \right) \left(
 {u - \frac{i}{g} (K+M) \over u +\frac{i}{g}(K+M)} \right)
\prod_{k=1}^{\text{min}(K,M)-1} \left( {u - \frac{i}{g} (|K-M|+2k) \over u +\frac{i}{g}(|K-M|+2k)} \right)^2
.\nn
\end{align}
The  elements $S^{\Sigma}_{\CQ'\CQ}$ are:
\eq
S^{\Sigma}_{\CQ'\CQ}(z,u)= (S_{\CQ' \CQ}(z-u))^{-1} (\Sigma_{\CQ'\CQ}(z,u))^{-2},
\en
where  $\Sigma_{\CQ'\CQ}$ is the improved dressing factor for the mirror bound states
\eq
\Sigma_{\CQ'\CQ}(z,u)=\prod_{k=1}^{\CQ'}\prod_{l=1}^{\CQ}\left(\frac{1-\frac{1}{x(z+\frac{i}{g}(\CQ'+2-2k))x(u+\frac{i}{g}(\CQ-2l))}}{1-\frac{1}{x(z+\frac{i}{g}(\CQ'-2k))
x(u+\frac{i}{g}(\CQ+2-2l))}}\right)\sigma_{\CQ' \CQ}(z,u),
\en
with $\sigma_{\CQ'\CQ}$ evaluated in the mirror kinematics. A precise analytic expression for the mirror improved dressing factor has been given in~\cite{Arutyunov:dressingfactor}, and a more compact integral representation in~\cite{Gromov:2009bc}. We show in Appendices~\ref{AB} and~\ref{AC} the equivalence of the two results.
%
%
%
%
%%%%%%%%%%%%%%%%%%%%%%%%%%%%%%%%%%%%%%%%%%%%%%%%%%%%%%%%%%%%%%%%%%%%%%%%%%%%%%%%%%%%%%%%
%
\resection{The kernels and their properties}
\label{Akernels}
The functions  $\phi_M$  and $\phi_{MN}$ are relativistic-like  kernels  common to many models,
from the  {\bf xxx} quantum spin chain to the Hubbard model:
\begin{align}
\label{aM}
\begin{split}
\phi_M(u)&= \frac{M/g}{\pi (u^2+(M/g)^2)},  \\
\phi_{MN}(u)&= \phi_{|M-N|}(u) + 2\phi_{|M-N|+2}(u) + \dots + 2 \phi_{M+N-2}(u) +\phi_{M+N}(u).
\end{split}
\end{align}
Their behavior is summarized by the following statement: $\phi_M (u)$ has two poles at $u = \pm iM/g $, with  residues $ \pm 1 $.
Consider, for example,  the quantity $\CG \ast \phi_M (u) \equiv \int_{\RR} dz \; \CG(z) \phi_M(z-u)$. In terms of the integration variable $z$, there are two poles at $ z=u \pm iM/g$ and,  by analytically continuing  $\CG \ast \phi_M (u)$ to $\CG \ast \phi_M (u \pm i/g)$, we find
\begin{align}
\label{eq:propkernela}
\begin{split}
 \CG \ast \phi_M(u + i/g)+ \CG \ast \phi_M(u - i/g )&= \CG \ast \phi_{M+1}(u)+ \CG \ast \phi_{M-1}(u)
+  \CG(u)\delta_{M, 1}  \\
&=\CG \ast \phi_{M,1}(u) + \CG(u)\delta_{M,1}.~~~~
\end{split}
\end{align}
A simple way  to understand  the result (\ref{eq:propkernela}) is by noticing  that the contour of integration can be freely shifted  slightly  below  the real axis. Then  for $M = 1$, while     $\CG \ast \phi_1 (u) \rightarrow \CG \ast \phi_1 (u+i/g)$  the pole of $\phi_1 (z-u)$ at  $z-u=-i/g$ crosses the contour of integration. The corresponding residue  is precisely the   local term  $\CG(u)$ in the rhs of (\ref{eq:propkernela}).

Similarly:
\eq
\CG \ast \phi_{MN} (u +i/g) + \CG \ast \phi_{MN}(u -i/g) = \sum_{K} \I_{M K} \; \CG \ast \phi_{K N}(u) + \I_{M N} \CG(u).
\en
Suppose that $\CG(u)$ is now a function with square root branch points at $u=-2$ and $u=2$, as for example $L^{(\alpha)}_y(u)$. For $u \in (-2, 2)$ we have:
\eq
\CG \so\phi_N(u+i/g)+ \CG\so\phi_N(u-i/g )
=\left(\CG_{-}(u) - \CG_{+}(u)\right)\delta_{N, 1} +  \CG \so\phi_{N+1}(u)+\CG\so\phi_{N-1}(u)
\label{sres}
\en
where  $\CG_{-}(u)$ and $\CG_{+}(u)$ indicate  the function $\CG$ evaluated respectively on the
first and second sheets.
Computing the same expression  for   $ u \notin (-2, 2)$ leads   to the result (\ref{sres})  without  the
term $\left(\CG_{-}(u) - \CG_{+}(u)\right)\delta_{N, 1}$. This simple argument reveals the presence of  a pair of branch points for
$ \CG\so\phi_N(u)$  at $ u=-2\pm iN/g$ and $u=2 \pm iN/g$.
The non-relativistic nature of the model is evident from  the elements of the scattering matrix
$S_{y, \CQ}(z,u)$ and  $S_{(v|M), \CQ}(z,u)$: they depend  separately on the variables  $z$ and $u$.
Because also these kernels have isolated poles with residues $\pm 1$, the analogue of the shifting properties reported above
can be computed in a similar fashion, but require some additional care due to the fact that the singularities can now disappear inside a branch cut instead of hitting the real line.
For example, for $u \in (-2, 2)$:
\eq
\CG*\phi_{\CQ,y}( u+i/g)+ \CG*\phi_{\CQ,y}( u-i/g)
= \CG*\phi_{\CQ,(v|1)}( u)+\delta_{\CQ,1} \CG(u).
\label{eq:nrev}
\en
The validity of  (\ref{eq:nrev}) is restricted to $u \in  (-2, 2)$, and a comment
must be made on a possible ambiguity. Obviously, the continuation outside this range is possible, and the rhs of  (\ref{eq:nrev}) keeps  the same form. However, as soon as $\RRe(u)$  leaves the interval $(-2, 2)$, on the lhs   a branch cut comes between $\phi_{1,(v|1)}(z, u-i/g)$  and  $\phi_{1,(v|1)}(z, u+i/g)$ and these two points are not anymore connected by  vertical analytic continuation.

%This anticipates a general phenomenon: the cuts of the Y functions have exactly the same separation $i2/g$ of the shifts needed to reduce the TBA equations to a local form, so they  cannot be avoided.
Other useful identities are listed below. For $u \in (-2, 2)$:
\begin{align}
\label{eq:nrev2}
\begin{split}
\CG*\phi_{\CQ,(v|N)}( u+i/g)+
\CG&*\phi_{\CQ,(v| N)}(u-i/g)=
 \sum_{M} \I_{NM} \; \CG*\phi_{\CQ,(v| M)}(u)  \\
&+\delta_{\CQ, N+1} \CG(u) +\delta_{N, 1} (\CG*\phi_{\CQ,(y|-)}(u) - \CG*\phi_{\CQ,(y|+)}(u)).
\end{split}
\end{align}
To write (\ref{eq:nrev2}) we  have  used the properties
\begin{align}
\begin{split}
\frac{1}{2 \pi i} \frac{d}{dz} \ln \biggl(\frac{x(z+i\CQ/g)}{x(z-i\CQ/g)} \; \frac{x(z -i\CQ/g) - x(u+i2/g)}{x(z +i\CQ/g ) - x(u + i2/g)}&\,\frac{x(z-i\CQ/g)-x(u-i2/g)}{x(z+i\CQ/g)-x(u-i2/g)} \biggr) \\
&=\phi_{\CQ,(v|2)}(z, u)-\phi_{\CQ}(z-u),
\end{split}
\label{factors}
\end{align}
\eq
\phi_Q(z-u)=\phi_{\CQ,(y|-)}(z, u)+\phi_{\CQ,(y|+)}(z, u),
\label{eq:sumofphis}
\en
where
\eq
\phi_{\CQ,(y|-)}(z, u) \equiv \phi_{\CQ,y}(z, u),~~  \phi_{\CQ,(y|+)}(z, u) \equiv \phi_{\CQ,y}(z, u_*).
\en
For $u \in \RR$ we  have:
\eq
\CG*\phi_{(v|N), \CQ}(u+i/g)+ \CG*\phi_{(v|N), \CQ}(u-i/g) = - \delta_{\CQ, N+1} \CG(u) + \sum_{\CQ'} \I_{\CQ'\CQ}  \; \CG*\phi_{(v|N), \CQ'}(u).
\label{eq:nrev3}
\en
The equation that expresses the effect of the shifts on the kernel $\phi_{y,\CQ}$ has some extra
subtlety:
\eq
\CG \so\phi_{y,\CQ}(u+i/g)+ \CG\so\phi_{y,\CQ}( u-i/g)
=-\delta_{\CQ, 1} \CG(u) + \sum_{\CQ'} \I_{\CQ' \CQ} \; \CG\so \phi_{y,\CQ'}(u).
\label{asy}
\en
Since $\CG(u) \equiv \CG_-(u)$, the property (\ref{asy})   is at the origin of the  asymmetry between $Y^{(\alpha)}_{(y|-)}(u) \equiv Y^{(\alpha)}_{y}(u)$   and
$Y^{(\alpha)}_{(y|+)}(u) \equiv Y^{(\alpha)}_{y}(u_*) $ on the rhs of (\ref{eq:YQf}).
Finally, another useful relation was proved in~\cite{Arutyunov:simplified}:
\eq
\CG*K^{\Sigma}_{\CQ' \CQ}(u+i/g)+\CG*K^{\Sigma}_{\CQ' \CQ}(u-i/g) = \sum_{P} \I_{\CQ P} \CG*K^{\Sigma}_{\CQ' P}(u),~(|\RRe(u)|<2).
\label{DF}
\en

%%%%%%%%%%%%%%%%%%%%%%%%%%%%%%%%%%%%%%%%%%%%%%%%%%%%%%%%%%%%%%%%%%
%
\resection{The  dressing factor in the direct and mirror theories }
\label{AB}
In Section~\ref{deltas}, the contribution to $\Delta$ coming from the improved dressing factor of the mirror theory
was  denoted by  $\Delta^{\Sigma}$:
\eq
\Delta^{\Sigma}(u) = 2\sum_{\CQ} L_{\CQ}\ast K_{\CQ}^{\Sigma}(u),
\en
with
\eq
K_{\CQ}^{\Sigma}(z,u)= K_{\CQ,1}^{\Sigma}(z,u+i/g)-K_{\CQ,1}^{\Sigma}(z,u_* +i/g),
~~
K_{\CQ,1}^{\Sigma}(z,u) = {1 \over 2 \pi i } {d \over dz} \ln \Sigma_{\CQ,1}(z,u),
\en
and in~\cite{Arutyunov:simplified}  the following expression was derived:
\begin{align}
\begin{split}
\Delta^{\Sigma}(u)=& 2\sum_{N}\int_{\mathbb{R}} dz \; \sum_{\CQ} L_{\CQ}(z)\left(K(z+i(2N+\CQ)/g, u)+K(z-i(2N+\CQ)/g, u)\right)  \\
&-2 \int_{\mathbb{R}} dz \; \sum_{\CQ} L_{\CQ}(z)\sum_{N} \oint_{\bgammao} ds \; \phi_{\CQ , y}(z, s)\left(K(s+i2N/g, u)+K(s-i2N/g, u)\right) \\
=& 2\sum_{N}\int_{\mathbb{R}} dz \; \sum_{\CQ} L_{\CQ}(u)\left(K(z+i(2N+\CQ)/g, u)+K(z-i(2N+\CQ)/g, u)\right)  \\
&+  \sum_{N,\alpha} \oint_{\bgammao} dz \;\ln Y_y^{(\alpha)}(z)\left(K(z+i2N/g, u)+K(z-i2N/g, u)\right).
\end{split}
\label{eq:deltasigma2}
\end{align}
The aim of this appendix is to  show how to recast (\ref{eq:deltasigma2}) into a form that emphasizes the symmetries between the direct and the mirror theory.
Let us start from the following  simple property. For any kernel $f(z, u)$ which is analytic in a neighborhood of the real axis, and for any function $r(z)$ with a square root branch cut along $(-\infty, -2) \cup (2, +\infty)$, we have:
\eq
\int_{\RR \pm i\epsilon} dz \;(r(z)-r(z_*))f(z, u)=\oint_{\bgammao}  dz \; r(z)f(z, u)\pm \oint_{\bgammax}  dz \; r(z)f(z, u).
\label{eq:property}
\en
Now, let us consider the following example:
\eq
\int_{\RR \pm i\epsilon}  dz \;\ln\left(Y_{(y|-)}^{(\alpha)}(z)/Y_{(y|+)}^{(\alpha)}(z)\right)f(z, u)= \ln Y_y^{(\alpha)} \so \; f( u) \pm \ln Y_y^{(\alpha)} \sx \;f( u).
\label{eq:example}
\en
In this case a further step is possible thanks to relation (\ref{dis1}). In fact, as long as the kernels $f(z, u)$ and $h(z, u)$ are analytic and sufficiently damped at infinity in the upper and lower half plane respectively, the lhs of (\ref{eq:example}) can be rewritten as a pair of dispersion relations:
\begin{align}
\begin{split}
\int_{\mathbb{R} + i\epsilon}  dz \;\ln\left(Y_{(y|-)}^{(\alpha)}(z)/Y_{(y|+)}^{(\alpha)}(z)\right)f(z, u)&=-\oint_{\Gamma_{\text{O}}^{+}}dz \;\ln\left(Y_{(y|-)}^{(\alpha)}(z)/Y_{(y|+)}^{(\alpha)}(z)\right)f(z, u);\\
\int_{\mathbb{R} - i\epsilon}  dz \;\ln\left(Y_{(y|-)}^{(\alpha)}(z)/Y_{(y|+)}^{(\alpha)}(z)\right)h(z, u)&=\oint_{\Gamma_{\text{O}}^{-}}dz \;\ln\left(Y_{(y|-)}^{(\alpha)}(z)/Y_{(y|+)}^{(\alpha)}(z)\right)h(z, u),
\end{split}
\end{align}
where the integration contours $\Gamma_{\text{O}}^{+}$, $\Gamma_{\text{O}}^-$ run inside the upper (or lower, respectively) half plane and, as the horizontal size of rectangles goes to infinity, $\Gamma_{\text{O}} \sim \Gamma_{\text{O}}^{+} \cup \Gamma_{\text{O}}^-$ ($\Gamma_{\text{O}}$ is represented in Figure~\ref{Gamma0}).
It is then sufficient to rephrase the derivation of Section~\ref{sectionDis} to transform (\ref{eq:example}) into:
\begin{align}
\begin{split}
\sum_{\CQ} \int_{\mathbb{R}}dz \;L_{\CQ}(z) f(z+i\CQ/g, u)&= \sum_{\alpha} \left(- \ln Y_y^{(\alpha)} \so \; f(u)- \ln Y_y^{(\alpha)} \sx \; f( u) \right); \\
\sum_{\CQ} \int_{\mathbb{R}} dz\;  L_{\CQ}(z) h(z-i\CQ/g, u)&=\sum_{\alpha} \left(-\ln Y_y^{(\alpha)} \so \; h(u)+ \ln Y_y^{(\alpha)} \sx \; h(u) \right).
\end{split}
\label{eq:property2}
\end{align}
These identities are appropriate to simplify (\ref{eq:deltasigma2}). In fact, taking $f(z, u) = \sum_{N=1}^{\infty} K(z+i2N/g, u)$ and $h(z, u) = \sum_{N=1}^{\infty} K(z-i2N/g, u)$ and summing the respective equations we find:
\begin{align}
\begin{split}
\int_{\mathbb{R}} dz \; \sum_{\CQ}L_{\CQ}(z)\sum_{N}&{\Big (}K(z+i(2N+\CQ)/g, u)+K(z-i(2N+\CQ)/g, u){\Big )}\\
=&-\sum_{\alpha}\oint_{\bgammao}dz \; \ln Y_y^{(\alpha)}(z)\sum_{N}{\Big (}K(z+i2N/g, u)+K(z-i2N/g, u){\Big )}\\
&-\sum_{\alpha}\oint_{\bgammax}dz \;\ln Y_y^{(\alpha)}(z)\sum_{N}{\Big (}K(z+i2N/g, u)-K(z-i2N/g, u){\Big )}.
\end{split}
\end{align}
Therefore, we arrive at the following  equation:
\eq
\Delta^{\Sigma}(u)=\sum_{N,\alpha} \oint_{\bgammax} dz \; \ln Y_y^{(\alpha)}(z)\left(-K(z+i2N/g, u)+K(z-i2N/g, u)\right).
\en
Introducing the function
\eq
K_{\Gamma}^{[N]}(z) = {1 \over 2 \pi i } {d \over dz} \ln {\Gamma(N/2-igz/2)  \over \Gamma(N/2+igz/2) },
\en
which grows at most as $\ln|z|$ as $|z|  \rightarrow \infty$ and
using the property  $K(z, u) \sim 1/z^2$ for $|z| \rightarrow \infty$  we can write
\eq
\oint_{\bgammax}dz \; K_{\Gamma}^{\left[2\right]} (s- z) K(z, u) =\sum_{N} \left(K(s+i2N/g, u)-K(s-i2N/g, u)\right)
+K_{\Gamma}^{\left[2\right]} (s- u).
\label{KK}
\en
Equation (\ref{KK}) and the discontinuous parts of the TBA equation (\ref{TBA2}) (independently re-derived in Section~\ref{sectionDis}  and Appendix \ref{AE}) lead to
\begin{align}
\Delta^{\Sigma}(u)
=&-\sum_{\alpha}\oint_{\bgammax} ds \; \ln Y_y^{(\alpha)}(s)\oint_{\bgammax}dz \; K_{\Gamma}^{\left[2\right]} (s- z) K(z, u)+\sum_{\alpha}\oint_{\bgammax} ds \; \ln Y_y^{(\alpha)}(s)K_{\Gamma}^{\left[2\right]} (s- u)\nn\\
=&2\sum_{\CQ'} L_{\CQ'}\ast\oint_{\bgammax} d\z\; \phi_{\CQ, y}(\z){\Big (} \oint_{\bgammax}d\zi \;  K_{\Gamma}^{\left[2\right]} (\z- \zi) K(\zi, u)-K_{\Gamma}^{\left[2\right]}(\z-u){\Big )},
\label{eq:deltasigma3}
\end{align}
with $u \notin $ $(-\infty, -2) \cup (2, +\infty)$.
Notice that the rhs of equation  (\ref{eq:deltasigma3}) coincides with  the discontinuity of  the function
\eq
2\sum_{\CQ'} L_{\CQ'}\ast\oint_{\bgammax} d\z \; \phi_{\CQ', y}(\z)\oint_{\bgammax}d\zi  \; K_{\Gamma}^{\left[2\right]}(\z-\zi)\frac{1}{2 \pi i} \frac{d}{d\zi}\ln(x(\zi)-x(u))
\label{eq:deltasubstitute}
\en
across  $(-\infty, -2) \cup (2, +\infty)$. Therefore we are free to trade (\ref{eq:deltasubstitute}) for
$\frac{1}{2}\Delta^{\Sigma}(u)$ every time we need to evaluate convolutions of the kind $\oint_{\bgammax} du \;\Delta^{\Sigma}(u) a(u, v) $. We encountered such an expression when we derived a dispersion relation for the $\CQ$-particles, see (\ref{eq:dressingrelated}). By property (\ref{eq:residue1}), we can now evaluate it to be:
\eq
\frac{1}{2}\; \Delta^{\Sigma} \sx \; \phi_{\CQ}( v)=-2\sum_{\CQ'} L_{\CQ'}\ast\oint_{\bgammax}
d\z \;\phi_{\CQ', y}(\z)\oint_{\bgammax}d\zi \; K_{\Gamma}^{\left[2\right]}(\z-\zi)\phi_{y, \CQ}(\zi, v).
\label{inte}
\en
This form resembles closely the integral representation of the direct theory dressing kernel found in~\cite{Dorey:2007xn} (see also~\cite{Arutyunov:dressingfactor}):
\begin{align}
\begin{split}
{\ln\sigma}_{\CQ' \CQ}(u,v)=& \chi(x_{\ts}(u+i\CQ'/g),x_{\ts}(v+i\CQ/g)) + \chi(x_{\ts}(u-i\CQ'/g),x_{\ts}(v-i\CQ/g)) \\
&-\chi(x_{\ts}(u+i\CQ'/g),x_{\ts}(v-i\CQ/g)) - \chi(x_{\ts}(u-i\CQ'/g),x_{\ts}(v+i\CQ/g)),
\end{split}
\label{dorey}
\end{align}
where
\eq
x_{\ts}(u)= \left( \frac{u}{2}  +  u\sqrt{ \frac{1}{4}- \frac{1}{u^2}} \right),~
\label{xsu}
\en
with $|x_{\ts}| \ge 1$ on  the first Riemann sheet. Notice that $x(u)$ and $x_{\ts}(u)$  are different sections of the
same function:  $x_{\ts}(u)$ has a cut on the segment  $u \in (-2,2)$ while  the branch cut of $x(u)$ is on
$(-\infty,-2) \cup (2,+\infty)$. More explicitly:
\eq
x_{\ts}(u) =  \left\{ \begin{array}{lll}
 x(u) & \hbox{for} & \IIm(u)<0;  \\
& & \\
1/x(u) & \hbox{for} & \IIm(u)>0.
\end{array}\right.
\label{xxs}
\en
In (\ref{dorey}) we have also introduced the function
\eq
\chi(a,b)= i \oint {d x_1 \over 2 \pi i} \oint {d x_2 \over 2 \pi i} { 1 \over (x_1 -a)}{ 1 \over (x_2 -b)}
\ln {\Gamma[1+ ig(x_1+1/x_1 -x_2-1/x_2 )/2]\over \Gamma[1- ig(x_1+1/x_1 -x_2-1/x_2  )/2] },
\en
where the integrals run  over the unit circles $|x_1|=|x_2|=1$. After a simple change of variables $x_1=x_{\ts}(\z)$,
$x_2=x_{\ts}(\zi)$ and an integration by parts we get
\eq
\frac{1}{2 \pi i}\frac{d}{du}{\ln\sigma}_{\CQ'\CQ}(u, v) =-\oint_{\bgammao} d\z \; \phi^{\ts}_{\CQ', y}(u, \z) \oint_{\bgammao} d\zi \; K_{\Gamma}^{\left[2\right]}(\z-\zi)\phi^{\ts}_{y, \CQ}(\zi, v).
\label{eq:hofman}
\en
In  (\ref{eq:hofman})
the superscripts indicate that the kernels $\phi^{\ts}_{\CQ, y}$, $\phi^{\ts}_{y, \CQ}$ differ from those defined in Appendix~\ref{AA} by the replacement $x \rightarrow x_{\ts}$\footnote{Notice  that  a  (zero-contribution)  extra factor    $\frac{1}{2 \pi i} \frac{d}{du}\ln \sqrt{\frac{x(u+i\CQ/g)}{x(u-i\CQ/g)}}$ has been included in (\ref{eq:hofman}) to emphasize
its  relationship  with  the kernels $\phi_{\CQ, y}$ and $\phi_{y,\CQ}$.}.
From (\ref{inte}) we see that (\ref{eq:dressingrelated}) is fulfilled iff the mirror theory improved dressing kernel admits
an integral representation analogous  to (\ref{eq:hofman}):
\eq
K_{\CQ'\CQ}^{\Sigma}(u,v)  = \frac{1}{2 \pi i}\frac{d}{du}{\ln\Sigma}_{\CQ' \CQ}(u, v)=\oint_{\bgammax} d\z \; \phi_{\CQ', y}(u, \z) \oint_{\bgammax} d\zi \; K_{\Gamma}^{\left[2\right]}(\z-\zi)\phi_{y, \CQ}(\zi, v).~~~~~~
\label{eq:hofman2}
\en
This is precisely  the result found in~\cite{Gromov:2009bc}.
More specifically, we just need (\ref{eq:hofman2}) to be valid  when the dressing kernel is convolved with $\sum_{\CQ'} L_{\CQ'}$. This concludes the derivation of the TBA equations for the $\CQ$-particle described in Section~\ref{Qnodes}. In the following appendix we have proved (\ref{eq:hofman2}) starting from the formula for $\Sigma_{\CQ' \CQ}$ given in~\cite{Arutyunov:dressingfactor}.\\

\noindent Notice that (\ref{eq:hofman2}) can be obtained from the Dorey-Hofman-Maldacena formula
(\ref{eq:hofman}) by the following three steps: switching to the mirror kinematics, changing all the contours of integration from $\bgammao$ to $\bgammax$, and imposing an overall minus sign. Therefore, by this duality transformation we have related the  original  dressing factor $\sigma_{\CQ' \CQ}$ in the $su(2)$ sector of the direct theory to the  $sl(2)$  improved dressing factor $\Sigma_{\CQ' \CQ}$ of the mirror theory.
%
%
%%%%%%%%%%%%%%%%%%%%%%%%%%%%%%%%%%%%%%%%%%%%%%%%%%%%%%%%%%%%%%%%%%%%%%%%%%%%%%%%%%%%%%%%
%
\resection{The improved dressing factor revised}
\label{AC}
The main objective of this section is to  prove  equation (\ref{eq:hofman2}) starting from the expression for the mirror improved dressing factor obtained in~\cite{Arutyunov:dressingfactor}.
The following identities  are useful to convert the  result of~\cite{Arutyunov:dressingfactor} in our notation:
\begin{align}
\begin{split}
\frac{d}{du} \Phi(x(u), x(v)) &= i\oint_{\bgammao} \frac{ds}{2 \pi i} \frac{d}{du}\ln(x(s)-x(u))\oint_{\bgammao} \frac{dt}{2 \pi i} K^{\left[2\right]}_{\Gamma}(s-t)\frac{d}{dt}\ln(x(t)-x(v)), \\
\frac{d}{du}\Psi(x(u), x(v))&= i\oint_{\bgammao}\frac{ds}{2 \pi i}\frac{d}{du}\ln(x(s)-x(v))K^{\left[2\right]}_{\Gamma}(u-s).
\end{split}
\end{align}
Then one can check that the mirror improved dressing kernel takes the form:
\begin{align}
\label{eq:dressingkernel}
K_{\CQ'\CQ}^{\Sigma}(u,v) =& -\oint_{\bgammao} d\z \; \phi_{\CQ', y}(u, \z) \oint_{\bgammao} d\zi\; K_{\Gamma}^{\left[2\right]}(\z-\zi)\phi_{y, \CQ}(\zi, v)  \nn \\
&-\frac{1}{2}\oint_{\bgammao} d\z \; \phi_{\CQ',y}(u, \z)\left(K_{\Gamma}^{\left[2 \right]}(\z-v+i\CQ/g)+K_{\Gamma}^{\left[2 \right]}(\z- v-i\CQ/g)\right) \nn \\
&+\frac{1}{2}\oint_{\bgammao} d\z \;\left(K_{\Gamma}^{\left[2 \right]}(u+i\CQ'/g- \z)+K_{\Gamma}^{\left[2 \right]}(u-i\CQ'/g-\z)\right)\phi_{y, \CQ}(\z, v)  \\
&+K_{\Gamma}^{\left[\CQ+\CQ'\right]}(u-v)+\frac{1}{2 \pi i}\frac{d}{du} \ln\frac{(1-\frac{1}{x(u+i\CQ'/g)x(v-i\CQ/g)})}{(1-\frac{1}{x(v+i\CQ/g)x(u-i\CQ'/g)})}\sqrt{\frac{x(u+i\CQ'/g)x(v-i\CQ/g)}{x(v+i\CQ/g)x(u-i\CQ'/g)}}.
\nn
\end{align}
Let us rearrange some terms in this formula. First, by using the identity
\eq
K^{\left[2\right]}_{\Gamma}(u+i\CQ/g)+K^{\left[2\right]}_{\Gamma}(u-i\CQ/g)= 2 K_{\Gamma}^{\left[\CQ+2\right]}(u)+\phi_{\CQ}(u)
\en
we can rewrite (\ref{eq:dressingkernel}) as:
\begin{align}
K_{\CQ'\CQ}^{\Sigma}(u,v) =&-\oint_{\bgammao} d\z \; \phi_{\CQ', y}(u, \z) \oint_{\bgammao} d\zi \; K_{\Gamma}^{\left[2\right]}(\z-\zi)\phi_{y, \CQ}(\zi, v) \nn \\
&-\oint_{\bgammao} d\z \; \phi_{\CQ',y}(u, \z)K_{\Gamma}^{\left[2 +\CQ\right]}(\z-v)-\frac{1}{2}\oint_{\bgammao} d\z \; \phi_{\CQ',y}(u, \z)\phi_{\CQ}(\z - v) \nn \\
&+\oint_{\bgammao} d\z \; K_{\Gamma}^{\left[2+\CQ'\right]}(u-\z)\phi_{y, \CQ}(\z, v)+\frac{1}{2}\oint_{\bgammao} d\z \; \phi_{\CQ'}(u-\z)\phi_{y, \CQ}(\z , v) \\
&+K_{\Gamma}^{\left[\CQ+\CQ'\right]}(u-v)+\frac{1}{2 \pi i}\frac{d}{du} \ln\frac{(1-\frac{1}{x(u+i\CQ'/g)x(v-i\CQ/g)})}{(1-\frac{1}{x(v+i\CQ/g)x(u-i\CQ'/g)})}\sqrt{\frac{x(u+i\CQ'/g)x(v-i\CQ/g)}{x(v+i\CQ/g)x(u-i\CQ'/g)}}.
\nn
\end{align}
Next, we notice that, subtracting the identities (\ref{eq:QQ'convol1}) and (\ref{eq:QQ'convol2}), we find
\begin{align}
\begin{split}
-\frac{1}{2}&\left(\oint_{\bgammao} d\z \; \phi_{\CQ',y}(u, \z)\phi_{\CQ}(\z - v)-\oint_{\bgammao} d\z \; \phi_{\CQ'}(u-\z)\phi_{y, \CQ}(\z , v)\right)\\
&=-\phi_{\CQ+\CQ'}(u-v)-\frac{1}{2 \pi i} \frac{d}{du}\ln\frac{(1-\frac{1}{x(u+i\CQ'/g)x(v-i\CQ/g)})}{(1-\frac{1}{x(v+i\CQ/g)x(u-i\CQ'/g)})}\sqrt{\frac{x(u+i\CQ'/g)x(v-i\CQ/g)}{x(v+i\CQ/g)x(u-i\CQ'/g)}},
\end{split}
\end{align}
so that the dressing kernel can be rewritten as:
\begin{align}
\label{eq:alternativedressing}
%\protect\begin{split}
K_{\CQ'\CQ}^{\Sigma}(u,v) =&-\oint_{\bgammao} d\z \; \phi_{\CQ', y}(u, \z) \oint_{\bgammao} d\zi \; K_{\Gamma}^{\left[2\right]}(\z-\zi)\phi_{y, \CQ}(\zi, v)
-\oint_{\bgammao} d\z \; \phi_{\CQ',y}(u, \z)K_{\Gamma}^{\left[2 +\CQ\right]}(\z-v) \nn \\
&+\oint_{\bgammao} d\z \; K_{\Gamma}^{\left[2+\CQ'\right]}(u-\z)\phi_{y, \CQ}(\z, v)
+K_{\Gamma}^{\left[\CQ+\CQ'+2\right]}(u-v).
%\end{split}
\end{align}
We want to show that, under convolution with $\sum_{\CQ'}L_{\CQ'}$, this is perfectly equivalent to (\ref{eq:hofman2}):
\begin{align}
\label{eq:goal}
\begin{split}
\sum_{\CQ'} L_{\CQ'} \ast K_{\CQ'\CQ}^{\Sigma}(v) &=\sum_{\CQ'} L_{\CQ'} \ast \oint_{\bgammax} d\z \; \phi_{\CQ',y}^{(\alpha)}(\z)\oint_{\bgammax} d\zi\; K_{\Gamma}^{\left[2\right]}(\z-\zi) \phi_{y, \CQ}(\zi, v)\\
&=-\oint_{\bgammax} d\z \; \ln Y_y^{(\alpha)}(\z)\oint_{\bgammax} K_{\Gamma}^{\left[2\right]}(\z-\zi) \phi_{y, \CQ}(\zi, v).
\end{split}
\end{align}
To begin, let us apply property (\ref{eq:property2}) to change the contour of integration under the `$\z$' variable in the last line. Taking into account that $\Gamma(1+i\frac{g}{2}u)$ is free of singularities for $u$ in the lower half plane we find:
\begin{align}
-&\oint_{\bgammax}\; d\z  \; \ln Y_y^{(\alpha)}(\z)\oint_{\bgammax} \frac{d\zi}{2 \pi i} \;\frac{d}{d\z}\ln\frac{\Gamma(1-i {g\over 2} (\z-\zi))}{\Gamma(1+i{g\over 2}(\z-\zi))}\phi_{y, \CQ}(\zi, v)  \nn\\
&=\oint_{\bgammao} d\z \; \ln Y_y^{(\alpha)}(\z)\oint_{\bgammax}\frac{d\zi}{2 \pi i} \;\frac{d}{d\z}\ln{(\Gamma(1-i{g\over 2}(\z-\zi)) \Gamma(1+i{g\over 2}(\z-\zi)))}\phi_{y, \CQ}(\zi, v)  \\
&~~+\sum_{\CQ'}\int_{\mathbb{R}} d\z \; L_{\CQ'}(\z)\oint_{\bgammax} \frac{d\zi}{2 \pi i} \;\frac{d}{d\z}(\ln{\Gamma(1-i{g\over 2}(\z+i\CQ'/g-\zi)) \Gamma(1+i{g\over 2}(\z-i\CQ'/g-\zi))})\phi_{y, \CQ}(\zi, v).~~\nn
\end{align}
The next step  is a change of integration contour under the `$\zi$' variable by means of (\ref{eq:property}). We arrive at:
\begin{align}
-\oint_{\bgammax} d\z \; \ln Y_y^{(\alpha)}(\z)\oint_{\bgammax} &d\zi K^{\left[2\right]}_{\Gamma}(\z- \zi)\phi_{y, \CQ}(\zi, v)\nn\\
=& \oint_{\bgammao} d\z \;\ln Y_y^{(\alpha)}(\z)\oint_{\bgammao}d\zi \; K^{\left[2\right]}_{\Gamma}(\z- \zi)\phi_{y, \CQ}(\zi, v)\nn\\
&+\oint_{\bgammao} d\z \;\ln Y_y^{(\alpha)}(\z)\frac{d}{d\z}{\Big (}-\int_{\mathbb{R}-i \epsilon} \frac{d\zi }{2 \pi i}\; \ln{\Gamma(1-i\frac{g}{2}(\z-\zi))}\nn\\
&+\int_{\mathbb{R}+i \epsilon} \frac{d\zi}{2 \pi i} \; \ln{\Gamma(1+i\frac{g}{2}(\z-\zi))}{\Big)}(\phi_{(y|-), \CQ}(\zi, v)-\phi_{(y|+), \CQ}(\zi, v))\\
&+\sum_{\CQ'}\int_{\mathbb{R}} d\z \; L_{\CQ'}(\z)\oint_{\bgammao} d\zi \;  K_{\Gamma}^{\left[2+\CQ'\right]}(\z-\zi)\phi_{y, \CQ}(\zi, v)\nn\\
&+\sum_{\CQ'}\int_{\mathbb{R}}d\z \; L_{\CQ'}(\z)\frac{d}{d\z}{\Big (}-\int_{\mathbb{R}-i \epsilon} \frac{d\zi}{2 \pi i} \;  \ln{\Gamma(1-i\frac{g}{2}(\z+i\CQ'/g-\zi))}\nn\\
&+\int_{\mathbb{R}+i \epsilon} \frac{d\zi}{2 \pi i} \; \ln{\Gamma(1+i\frac{g}{2}(\z-i\CQ'/g-\zi))}{\Big)}(\phi_{(y|-), \CQ}(\zi, v)-\phi_{(y|+), \CQ}(\zi, v)).\nn
\end{align}
Using Jordan's lemma, one can easily evaluate the integrals over $\RR \pm i \epsilon$. Closing the contours of integrations with two semicircles of infinite radius in the upper or lower half plane respectively, we find:
\begin{align}
\frac{d}{d\z}{\Big (}-\int_{\mathbb{R}-i \epsilon} \frac{d\zi}{2 \pi i}  \ln\Gamma(1 -& i\frac{g}{2}(\z-\zi))+
\int_{\mathbb{R}+i \epsilon} \frac{d\zi}{2 \pi i} \; \ln{\Gamma(1+i\frac{g}{2}(\z-\zi))}{\Big)}\\
\times& (\phi_{(y|-), \CQ}(\zi, v)-\phi_{(y|+), \CQ}(\zi, v))=K_{\Gamma}^{\left[2+\CQ\right]}(\z-\zi);
 \nn
\end{align}
\begin{align}
\frac{d}{d\z}{\Big (}-\int_{\mathbb{R}-i \epsilon} \frac{d\zi}{2 \pi i}  \ln\Gamma(1 - i\frac{g}{2}(\z+&i\CQ'/g-\zi))+\int_{\mathbb{R}+i \epsilon} \frac{d\zi}{2 \pi i} \; \ln{\Gamma(1+i\frac{g}{2}(\z-i\CQ'/g-\zi))}{\Big)}\\
&\times (\phi_{(y|-), \CQ}(\zi, v)-\phi_{(y|+), \CQ}(\zi, v))=K_{\Gamma}^{\left[2+\CQ+\CQ'\right]}(\z-\zi), \nn
\end{align}
and this finally proves (\ref{eq:goal}).
%%%%%%%%%%%%%%%%%%%%%%%%%%%%%%%%
%
\resection{Some useful identities}
\label{AD}
In this appendix we collect some useful identities which fill the gap between (\ref{eqeq}) and (\ref{eq:cauchyQ}) and the standard form of the TBA equations. All the needed manipulations are straightforward applications of the following lemmas.

Consider a function $f(u)$ with a square root branch cut along $(-\infty, -2) \cup (2, +\infty)$. Let $f(u)$ have isolated poles with residues $+1$ at $u=u_1,\dots, u_n$ on the first sheet and at $u_*=v_1, \dots, v_m$ on the second sheet, but otherwise let it be analytic and asymptotically $o(u)$. Then a simple application of the residue theorem yields:
\eq
f \sx \; \phi_{\CQ}(u)=f(u+i\frac{\CQ}{g})-f(u-i\frac{\CQ}{g})+\sum_{i=1}^n \phi_{\CQ}(u_i-u),
\label{eq:residue1}
\en
 and
\eq
f \so \; \phi_{\CQ}(u)=f(u+i\frac{\CQ}{g})-f((u-i\frac{\CQ}{g})_*)+\sum_{i\text{}/\text{ }\IIm(u_i)>0}\phi_{\CQ}(u_i-u)+\sum_{i\text{}/\text{ } \IIm(v_i)<0}\phi_{\CQ}(v_i-u).\label{eq:residue1*}
\en
Now suppose that $g(u)$ is a generic function and that, for $u$ in the lower half plane, $f[g](u)$ is defined by the following integral representation:
\eq
\label{eq:definition}
f[g](u) = \oint_{\bgammao} \frac{dz}{2 \pi i} g(z) \frac{d}{dz}\ln (x(z)-x(u)),~~   (\IIm(u)<0).
\en
Notice that $ f[g](u)$ is still analytic in the specified range $\IIm(u)<0$ but that, when u is continued through $(-2, 2)$, (\ref{eq:definition}) needs to be replaced by another representation:
\eq
f[g](u)= \oint_{\bgammao} \frac{dz}{2 \pi i} g(z) \frac{d}{dz}\ln (x(z)-x(u)) -g(u),~~(\IIm(u)>0),
\en
showing that the analytic behaviour of $f[g](u)$ in the upper half plane depends on the specific form of $g(u)$ and equation (\ref{eq:residue1}) is in general incorrect.\\
However, the following relation is true for any $g(u)$:
\eq
\label{eq:residue2}
\oint_{\bgammax}dz \; f[g](z) \phi_{\CQ}(z-u)=-\oint_{\bgammao} dz \; g(z) \phi_{y, \CQ}(z, u)-\int_{\mathbb{ R}+i\epsilon} dz \;g(z) \phi_{\CQ}(z-u).
\en
As a first example, let us consider the integral:
\eq
\oint_{\bgammao} dz \; \ln Y_y^{(\alpha)}(z) \phi_M(z-u) =-\sum_{\CQ} \int_{\mathbb{R}} dz \; L_{\CQ}(z) {\Big(}\oint_{\bgammao} ds \; \phi_{\CQ, y}(z, s) \phi_{M}(s-u){\Big)},
\en
which we encounter in formula (\ref{eqeq}). Taking into account the position of the poles of $\phi_{\CQ, y}$ and using (\ref{eq:residue1*}) we find:
\begin{align}
\label{eq:QQ'convol1}
\oint_{\bgammao} &ds \; \phi_{\CQ, y}(z, s) \phi_M(s-u)\nn\\
=&\frac{1}{2 \pi i}\frac{d}{dz}\ln\frac{(x(z-i\CQ/g)-x(u+iM/g))(x(z+i\CQ/g)-\frac{1}{x(u-iM/g)})}{(x(z+i\CQ/g)-x(u+iM/g))(x(z-i\CQ/g)-\frac{1}{x(u-iM/g)})} + \phi_M(z-u+i\CQ/g) \nn\\
=&\phi_{\CQ, (v|M)}(z, u)-\phi_{\CQ}(z-u+iM/g)-\sum_{J=1}^{M-1} \phi_{\CQ-M+2J}(z-u)+\phi_M(z-u+i\CQ/g),
\end{align}
and, finally
\eq
 \ln Y^{(\alpha)}_y \so \; \phi_M(u) =-\sum_{\CQ} \left( L_{\CQ}*\phi_{\CQ, (v|M)}( u)-\sum_{I=0}^{M-1} L_{\CQ}*\phi_{\CQ-M+2I}(u) \right),
\label{eq:vintermediate}
\en
which  shows that (\ref{eqeq}) is the TBA equation for the $v$-particles. We report also the very similar result:
\begin{align}
\label{eq:QQ'convol2}
\begin{split}
\oint_{\bgammao} ds \; \phi_{M}(&u-s)\phi_{y, \CQ}(s , v)= -\phi_{M}(u-v -i\CQ/g)  \\
&+\frac{1}{ 2\pi i}\frac{d}{du}\ln\frac{(\frac{1}{x(u-iM/g)}-x(v+i\CQ/g)) (x(u+iM/g)-x(v-i\CQ/g))}{(\frac{1}{x(u-iM/g)}-x(v-i\CQ/g))(x(u+iM/g)-x(v+i\CQ/g))}.
\end{split}
\end{align}
Now let us deal with the convolution
\eq
\oint_{\bgammax} dz \; \ln Y_1(z+\frac{i}{g})\phi_{\CQ}(z-u).
\en
As a first step we notice that, inserting (\ref{eq:deltatba}), the latter can be broken up in the following terms:
\begin{align}
\oint_{\bgammax} dz \ln Y_1(z+\frac{i}{g})&\phi_{\CQ}(z-u)=\frac{1}{2}\oint_{\bgammax} dz \; \Delta(z)\phi_{\CQ}(z-u)\nn\\
=&\oint_{\bgammax}{\Big (} \int_{\mathbb{R}}\frac{dt}{2 \pi i} \sum_{N,\alpha} L_{(v|N)}^{(\alpha)}(t) \frac{d}{dt}\ln(x(t+iN/g)-x(z))(x(t-iN/g)-x(z))\nn\\
&+    L \ln x(z) + f[g_y](z)+\frac{1}{2}\Delta^{\Sigma}(z) {\Big )}\phi_{\CQ}(z-u)dz,
\label{eq:terms}
\end{align}
with $g_y(u)\equiv \sum_{\alpha} L_y^{(\alpha)}(u)$. Now, as an application of (\ref{eq:residue1}) we have, in the first place:
\eq\label{energy}
-\int_{\bgammax} dz \; \ln{x(z)} \phi_{\CQ}(z-u)=\ln\frac{x(u-\frac{i \CQ}{g})}{x(u+\frac{i \CQ}{g})}\equiv \widetilde{E}_{\CQ}(u),
\en
which shows the appearance of the required energy term. Another easy computation shows that
\begin{align}
&-\int_{\bgammax} \frac{dz}{2 \pi i} \frac{d}{dt}\ln\left(x(t+iN/g)-x(z)\right)\left(x(t-iN/g)-x(z)\right)\phi_{\CQ}(z-u) \nn\\
&=\frac{1}{2 \pi i} \frac{d}{dt}\ln\frac{(t+i\frac{N}{g}-u-i\frac{\CQ}{g})(t-i\frac{N}{g}-u-i\frac{\CQ}{g})}{(t+i\frac{N}{g}-u+i\frac{\CQ}{g})(t-i\frac{N}{g}-u+i\frac{\CQ}{g})}\frac{\left(x(t+i\frac{N}{g})-x(u-i\frac{\CQ}{g})\right)\left(x(t-i\frac{N}{g})-x(u-i\frac{\CQ}{g})\right)}{\left(x(t+i\frac{N}{g})-x(u+i\frac{\CQ}{g})\right)\left(x(t-i\frac{N}{g})-x(u+i\frac{\CQ}{g})\right)}\nn\\
&=\sum_{J=0}^{N}\phi_{\CQ-N+2J}(t-u)+\phi_{(v|N), \CQ}(t, u)=\sum_{J=1}^{\CQ}\phi_{N-\CQ+2J}(t-u)+\phi_{(v|N), \CQ}(t,u).
\end{align}
Finally, by means of the second lemma (\ref{eq:residue2}) we find:
\eq
-\int_{\bgammax} dz \; f[g_y](z)\phi_{\CQ}(z-u)=\int_{\mathbb{R}+i\epsilon}dz \; L_y^{(\alpha)}(z)\phi_{\CQ}(z-u)+\oint_{\bgammao} dz \; L_y^{(\alpha)}(z)\phi_{y, \CQ}(z, u).
\en
Using also the fundamental results (\ref{inte},\ref{eq:hofman2})  for the dressing-related part, we have come to the following:
\begin{align}
\begin{split}
-\oint_{\bgammax} &dz \; \ln Y_1(z+\frac{i}{g})\phi_{\CQ}(z-u)
=2\sum_{\CQ'}L_{\CQ'} \ast K^{\Sigma}_{\CQ' \CQ}(u)+L\widetilde{E}_{\CQ}(u) +\sum_{\alpha}  L_y^{(\alpha)}\so \; \phi_{y, \CQ}(u)  \\
&+\sum_{\alpha}\int_{\mathbb{R}+i\epsilon} dz \; L_y^{(\alpha)}(z)\phi_{\CQ}(z-u)
+\sum_{N,\alpha}  L_{(v|N)}^{(\alpha)}*{\Big(}\sum_{J=1}^{\CQ}\phi_{N-\CQ+2J}+\phi_{(v|N), \CQ} {\Big )(u)}.
\end{split}
\label{eq:Y1convolution}
\end{align}
The latter equation shows that the Cauchy dispersion relation (\ref{eq:cauchyQ}) coincides with the TBA equation (\ref{TBA1}).
%
%%%%%%%%%%%%%%%%%%%%%%%%%%%%%%%%%%%%%%%%%%%%%%%%%%%%%%%%%%%%%%%%%%
%
\resection{Completing the TBA equations for the fermionic nodes}
\label{AE}
Let us show how to reproduce the missing part of the TBA equations (\ref{TBA2}) and (\ref{eq:halfTBA}) using only the functional relations (\ref{dis1}) and the Y-system equations (\ref{yw},\ref{yv},\ref{Yysystem}).

First of all, we notice that (\ref{Yysystem}) implies the following constraint on the discontinuities of $y$-related functions:
\begin{align}
&\left[\ln\left({Y_{(y|-)}^{(\alpha)}  Y_{(y|+)}^{(\alpha)}} \right)+\ln\left({Y_{(y|-)}^{(\alpha)}/ Y_{(y|+)}^{(\alpha)}}\right)\right]_{2N}+\left[\ln \left({Y_{(y|-)}^{(\alpha)}Y_{(y|+)}^{(\alpha)}}\right)+\ln\left({Y_{(y|-)}^{(\alpha)} / Y_{(y|+)}^{(\alpha)}}\right)\right]_{2N-2}\nn\\
&~~~=2\left[ \La^{(\alpha)}_{(v|1)}-\La^{(\alpha)}_{(w|1)}- L_{1}\right]_{2N-1}=2\left[L^{(\alpha)}_{(v|1)}-L^{(\alpha)}_{(w|1)}+\ln\left({Y^{(\alpha)}_{(v|1)} /Y^{(\alpha)}_{(w|1)}}\right)-L_{1} \right]_{2N-1}.
\end{align}
($N \in \mathbb{Z}$).
Next we use the following identity, which is a consequence of the Y-system equations (\ref{yw}) and (\ref{yv}):
\begin{align}
\begin{split}
\left[\ln\left({Y^{(\alpha)}_{(v|1)} /Y^{(\alpha)}_{(w|1)}}\right)\right]_{\pm (2N-1)}&=\left[\ln\left({Y_{(y|-)}^{(\alpha)} / Y_{(y|+)}^{(\alpha)}}\right)\right]_{\pm(2N-2)}-\sum_{\CQ =2}^{N}\left[L_{\CQ}\right]_{\pm (2N-\CQ)}\\
+\sum_{J=1}^{N-1}\sum_{J'}&\I_{J J'}\left[L^{(\alpha)}_{(v|J)}-L^{(\alpha)}_{(w|J)}\right]_{\pm(2N-1-J')}+\left[L^{(\alpha)}_{(v|N)}-L^{(\alpha)}_{(w|N)}\right]_{\pm N},
\end{split}
\end{align}
with $N=1, 2, \dots$. Using  the identity (\ref{dis1}):
\eq
\left[\ln \left(Y^{(\alpha)}_{(y|-)} /Y^{(\alpha)}_{(y|+)} \right) \right]_{ \pm 2N}=- \sum_{\CQ=1}^N\left[L_{\CQ}\right]_{\pm (2N-\CQ)},
\en
we find:
\begin{align}
\left[\ln \left({Y_{(y|-)}^{(\alpha)} Y_{(y|+)}^{(\alpha)}}\right)\right]_{2N}+&\left[\ln\left({Y_{(y|-)}^{(\alpha)} Y_{(y|+)}^{(\alpha)}}\right)\right]_{2N-2}\nn\\
=&2\sum_{J=1}^{N}\left[L^{(\alpha)}_{(v|J)}-L^{(\alpha)}_{(w|J)}\right]_{\pm(2N-J)}+2\sum_{J=1}^{N-1}\left[L^{(\alpha)}_{(v|J)}-L^{(\alpha)}_{(w|J)}\right]_{\pm(2N-2-J)}\nn\\
&-\sum_{\CQ =1}^{N}\left[L_{\CQ} \right]_{\pm (2N-\CQ)}-\sum_{\CQ =1}^{N-1} \left[L_{\CQ}\right]_{\pm (2N-2-\CQ)}.
\end{align}
Using the fact that $\ln \left({Y_{(y|-)}^{(\alpha)} Y_{(y|+)}^{(\alpha)}}\right)$ is regular on the real line we can solve the previous equation and we finally get to:
\eq
\left[\ln \left({Y_{(y|-)}^{(\alpha)}  Y_{(y|+)}^{(\alpha)}} \right)\right]_{\pm 2N}=2\sum_{J=1}^{N}\left[L^{(\alpha)}_{(v|J)}-L^{(\alpha)}_{(w|J)}\right]_{\pm(2N-J)}-\sum_{\CQ=1}^N \left[L_{\CQ} \right]_{\pm (2N-\CQ)}.
\en

\resection{A Rosetta Stone for $\Ad$}
\label{AF}
The system of TBA equations arising from $\Ad$  is intricate enough that one might find a bit nonsense that three different notations are actually being used in~\cite{Bombardelli:2009ns, Gromov:2009bc, Arutyunov:2009ur}. Sharing our part of responsibility for this proliferation of languages, we would like to remedy by providing the reader with some clues to help in the translation.
\\

\noindent First of all, there are little differences in the definition of the string coupling, which is denoted $g$ by all the authors. In this paper, $g$ is related to the 't Hooft parameter $\lambda$ by $\lambda = 4 \pi^2 g^2$, and agrees with the $g$ introduced by Arutyunov, Frolov (AF) in \cite{Arutyunov:2009ur}. There is a factor of 2 between this definition and the one adopted by Gromov, Kazakov, Kozak and Vieira (GKKV) in \cite{Gromov:2009tv}. Therefore, in the rest of this section we will represent the coupling $g$ of \cite{Gromov:2009tv} using its  Roman Latex style variant:
$\gG \equiv g^{\text{Ref\cite{Gromov:2009tv}}}$.

\begin{table}[h]
\begin{center}
\begin{tabular}{|c|c|c|}
\hline
This paper & AF & GKKV\\
\hline
\hline
$g$     &     $g$     &  $2 \gG$ \\
\hline
\end{tabular}
\caption{\small The coupling constants.}
\label{tableF1}
\end{center}
\end{table}
\noindent Our definition of Y functions is conventional, in that $Y_a=e^{\eps_a}$, where $\eps_a$ is the ratio between the densities of holes and particles of species $a$. Other choices are possible. In particular, we remind the reader that the double index appearing in the notation of GKKV runs over the nodes of the diagram depicted in Figure~\ref{N4LN},
allowing for a compact expression  of  the Y-system (\ref{eq:YQf}-\ref{yv}):
\[\frac{Y_{a, s}(v+i/2) Y_{a, s}(v-i/2)}{Y_{a+1, s}(v) Y_{a-1, s}(v)}=\frac{(1+Y_{a, s+1}(v))(1+Y_{a, s-1}(v))}{(1+Y_{a+1, s}(v))(1+Y_{a-1, s}(v))},
\]
and that the $Y_{a, s}$ functions are related through the transformation
\[
Y_{a, s}(v)=\frac{T_{a, s+1}(v) T_{a, s-1}(v)}{T_{a+1, s}(v) T_{a-1, s}(v)}
\]
to a set of T functions obeying the Hirota bilinear equation (T-system): $$T_{a, s}(v+i/2) T_{a, s}(v-i/2)=T_{a+1, s}(v) T_{a-1, s}(v)+T_{a, s+1}(v) T_{a, s-1}(v).$$
The same authors have also introduced a more pictorial notation, through the identifications:
\bea
Y_{1, 1s}(v)&\equiv& Y_{\figym s}(v), \nn\\
Y_{2, 2s}(v)&\equiv& Y_{\figyp s}(v), \nn\\
Y_{1, sM}(v)&\equiv& Y_{\figw {s M}}(v), \nn\\
Y_{M, s}(v)&\equiv& Y_{\figv {s M}}(v), \nn\\
Y_{M, 0}(v)&\equiv& Y_{\figQ {M}}(v),\nn
\eea
with $s=\pm 1$, $M=1, 2, \dots$.

\noindent Moreover, GKKV use rescaled variables, following from a different definition of the function $x$ which expresses the Bethe roots in terms of the rapidities \footnote{In fact, compare the  definition (\ref{xu}) with the following one,  adopted in \cite{Gromov:2009tv}:
\[
x^{\text{Ref\cite{Gromov:2009tv}}} (v)+\frac{1}{x^{\text{Ref\cite{Gromov:2009tv}}}(v)}=\frac{v}{\gG}.
\]
}.
The notation of AF is much more similar to ours, the main difference being the inversion of the Y functions for the $\CQ$-particles.
The precise relation between Y functions in the different settings is shown in Table \ref{tableF2}.
\begin{table}[h]
\begin{center}
\begin{tabular}{|cc|cc|cc|}
\hline
This paper && AF && GKKV&\\
\hline
\hline
$Y^{(\alpha)}_{(y|-)}(u)$  &   { $\alpha=1, 2$}    &     $Y^{(\alpha)}_{-}(u)$  &      & $1/Y_{\figym s}({u \gG})$  &   { $s=-3+2\alpha=-1, +1$} \\
\hline
$Y^{(\alpha)}_{(y|+)}(u)$  &   { $\alpha=1, 2$}
    &     $Y^{(\alpha)}_{+}(u)$  &      & $Y_{\figyp s}({u \gG})$  &
    { $s=-3+2\alpha=-1, +1$}
    \\
\hline
$ Y^{(\alpha)}_{(w|M)}(u)$  &   { $\alpha=1, 2$},
$M=1, 2, \dots$    &  $Y^{(\alpha)}_{M|w}(u)$  &&  $Y_{\figw {s (M+1)}}({u  \gG})$ &
{ $s=-3+2\alpha=-1, +1$}
     \\
\hline
$ Y^{(\alpha)}_{(v|M)}(u)$  &   { $\alpha=1, 2$, $M=1, 2, \dots$}
   & $Y^{(\alpha)}_{M|vw}(u)$   &&  $1/Y_{\figv {s (M+1)}}({u  \gG})$  &
   {$s=-3+2\alpha=-1, +1$}
   \\
\hline
$ Y_{M}(u)$  &{ $M=1, 2, \dots$}
  &   $1/Y_{M}(u)$  &&     $1/Y_{\figQ { M}}({u  \gG})$ &\\
\hline
\end{tabular}
\caption{\small The Y functions.}
\label{tableF2}
\end{center}
\end{table}
In Table \ref{tableF3} we summarize the relations between our kernels and those defined by other authors. Notice that the two variables are swapped in the notation of GKKV, and hence that there is a relation between right convolutions in one system of notations and left convolutions in the other.
\begin{table}[h]
\begin{center}
\begin{tabular}{|cc|cc|cc|}
\hline
This paper && AF && GKKV&\\
\hline
\hline
$\phi_{(y|-),M}(u, z)$    &&     $-K_-^{y M}(u, z)$  &&   $-{\gG}{\cal R}^{10}_{M1}({z}{\gG}, {u}{\gG})$ & \\
\hline
$\phi_{(y|+), M}(u, z)$    &&     $K_+^{y M}(u, z)$  &&   $-\gG{\cal B}^{10}_{M1}({z}{\gG}, {u}{\gG})$ &\\
\hline
$\phi_{M, (y|-)}(u, z)$    &&     $K_-^{M y}(u, z)$  &&   ${\gG}{\cal R}^{01}_{1M}({z}{\gG}, {u}{\gG})$ & \\
\hline
$\phi_{M, (y|+)}(u, z)$    &&     $K_+^{M y}(u, z)$  &&   $\gG{\cal B}^{01}_{1M}({z}{\gG}, {u}{\gG})$ &\\
\hline
$\phi_M(u-z)$    &&     $K_M(u-z)$  &&  ${\gG}K_M({z}{\gG}-{u}{\gG})$ & \\
\hline
$ \phi_{MN}(u-z)$    &&  $K_{MN}(u-z)$  &&  ${\gG}K_{NM}({z}{\gG}-{u}{\gG})$  & \\
\hline
$\phi_{(v|M), \CQ}(u,z)$    &&   $-K_{vwx}^{M \CQ}(u, z)$   &&  $-{\gG}{\cal B}^{10}_{\CQ, M-2 }({z}{\gG}, {u}{\gG})-{\gG}{\cal R}^{10}_{\CQ M}({z}{\gG}, {u}{\gG})$ &\\
\hline
$\phi_{\CQ, (v|M)}(u, z)$    && $K^{\CQ M}_{xv}(u, z)$   &&  ${\gG}{\cal B}^{01}_{M-2, \CQ}({z}{\gG}, {u}{\gG})+{\gG}{\cal R}^{01}_{M \CQ}({z}{\gG}, {u}{\gG}) $  & \\
\hline
$\frac{1}{2 \pi i}\frac{d}{du}\ln \sigma_{MN}(u, z)$  &&  $\frac{1}{2 \pi i}\frac{d}{du}\ln \sigma_{MN}(u, z)$    &&   $-{\gG}{\cal S}_{NM}({z}{\gG}, {u}{\gG})$&  \\
\hline
\end{tabular}
\caption{\small The TBA kernels.	}
\label{tableF3}
\end{center}
\end{table}
In the last row, $\sigma_{MN}(u, z)$ (corresponding  to $\sigma_{MN}(\gG u, \gG z)$ in \cite{Gromov:2009tv}) is the dressing factor for the scattering of an $M$ and an $N$ particle in the direct theory, which is understood to be analytically continued to the mirror kinematics in the context of mirror TBA.\\
\noindent For the mirror improved dressing factor we have adhered to the notation of AF:
\eq
\Sigma_{M N}(u, z)=\prod_{k=1}^{M}\prod_{l=1}^{N}\left(\frac{1-\frac{1}{x(u+\frac{i}{g}(M+2-2k))x(z+\frac{i}{g}(N-2l))}}{1-\frac{1}{x(u+\frac{i}{g}(M-2k))
x(z+\frac{i}{g}(N+2-2l))}}\right)\sigma_{M N}(u, z).\nn
\en
On the other hand, the double product factor on the right hand side emerges from the kernels of GKKV through the following identity:
\begin{align}
\begin{split}
-{\gG}K_{NM}(z{\gG}-u &{\gG})+ \gG{\cal R}_{NM}^{(11)}(z \gG, u \gG)- {\gG}{\cal B}_{NM}^{(11)}(z \gG, u \gG )  \\
&=
\frac{1}{2 \pi i}\frac{d}{du}\ln\prod_{k=1}^{M}\prod_{l=1}^{N}\left(\frac{1-\frac{1}{x(u+\frac{i}{g}(M+2-2k))x(z+\frac{i}{g}(N-2l))}}{1-\frac{1}{x(u+\frac{i}{g}(M-2k))
x(z+\frac{i}{g}(N+2-2l))}}\right)^2\frac{x(u+\frac{i}{g}(M+2-2k))}{x(u+\frac{i}{g}(M-2k))}
\\
& =
\frac{1}{2 \pi i}\frac{d}{du}\ln\prod_{k=1}^{M}\prod_{l=1}^{N}\left(\frac{1-\frac{1}{x(u+\frac{i}{g}(M+2-2k))x(z+\frac{i}{g}(N-2l))}}{1-\frac{1}{x(u+\frac{i}{g}(M-2k))
x(z+\frac{i}{g}(N+2-2l))}}\right)^2-N{  \tilde p^M(u)}.
\end{split}
\end{align}
The only difference between the TBA equations as written in \cite{Gromov:2009tv} and in \cite{Arutyunov:2009ur, Bombardelli:2009ns} is originated by the last term of the preceding equality. As we have anticipated in Section \ref{dilog}, it results in the introduction of a chemical potential proportional to the total momentum of the state under consideration. By the trace condition, we expect the extra chemical potentials to be integer multiples of $2 \pi i$. Therefore, beside possibly the swapping of some pairs of excited levels, the substance of the analysis should be unaffected, and the TBA equations of \cite{Arutyunov:2009ur, Bombardelli:2009ns ,Gromov:2009tv} are an equivalent starting point to study the spectrum of $\Ad$.
%
%%%%%%%%%%%%%%%%%%%%%%%%%%%%%%%%%%%%%%%%%%%%%%%%%%%%%%%%%%%%%%%%%%%%%%%%%%%%%

\end{document}